\documentclass[onecolumn]{aastex6}

\pdfoutput=1 

\usepackage{graphicx}
\usepackage{natbib}
\usepackage{amsmath,amssymb}
\usepackage{xcolor}
\usepackage{pifont}
\usepackage{hyperref}

\newcommand{\equ}[1]{\eq~(\ref{equ:#1})} \newcommand{\eq}{Eq.} \newcommand{\eqs}{Eqs.} 
\newcommand{\ie}{{\it i.e.}}  \newcommand{\eg}{{\it e.g.}}  \newcommand{\cf}{{\it cf.}}
\newcommand{\etc}{{\it etc.}}
\newcommand{\Fig}{Fig.}   \newcommand{\Ref}{Ref.}
\newcommand{\Refs}{Refs.} \newcommand{\Sec}{Sec.}   
\newcommand{\Tab}{Tab.} 

\def\jcap{Jour. Cosmology and Astro-Particle Phys.} \def\mnras{M.N.R.A.S} \def\apj{Astrophys.J.} \def\apjl{Astrophys.J.Lett.} \def\apjs{Astrophys.J.Supp.}           \def\prd{Phys.Rev.D}  \def\aap{Astron.Astrophys.}    

\begin{document}

\shorttitle{On the direct correlation between gamma-rays and PeV neutrinos from blazars}
\shortauthors{S. Gao et al.}
\title{On the direct correlation between gamma-rays and PeV neutrinos from blazars}

\author{Shan Gao\altaffilmark{1}, Martin Pohl\altaffilmark{1,2}, Walter Winter\altaffilmark{1}}
\altaffiltext{1}{Deutsches Elektronen-Synchrotron (DESY), Platanenallee 6, D-15738 Zeuthen, Germany}
\altaffiltext{2}{Institute of Physics and Astronomy, University of Potsdam, D-14476 Potsdam, Germany}
\email{shan.gao@desy.de}

\begin{abstract}
We study the frequently used assumption in multi-messenger astrophysics that the gamma-ray and neutrino fluxes are directly connected because they are assumed to be produced by the same photohadronic production chain. An interesting candidate source for this test is the flat-spectrum radio quasar PKS B1424-418, which recently called attention of a potential correlation between an IceCube PeV-neutrino event and its burst phase.  We simulate both the multi-waveband photon and the neutrino emission from this source using a self-consistent radiation model.  We demonstrate that a simple hadronic model cannot adequately describe the spectral energy distribution for this source, but a lepto-hadronic model with sub-dominant hadronic component can reproduce the multi-waveband photon spectrum observed during various activity phases of the blazar. As a conclusion, up to about 0.3 neutrino events may coincide with the burst, which implies that the leptonic contribution dominates in the relevant energy band. We also demonstrate that the time-wise correlation between the neutrino event and burst phase is weak.
\end{abstract}

\keywords{}

\section{Introduction}
\label{sec_intro}

The diffuse flux of TeV-PeV neutrinos detected with IceCube indicates extraterrestrial neutrino emission from cosmic accelerators \citep{IceCube13,ICPEV} of yet unknown nature. Among the prime source candidates are blazars, active galactic nuclei (AGN) featuring a relativistic jet roughly oriented along the line of sight to the observer. There is a rich literature modeling the neutrino emission from blazars and searching for positional and temporal correlations between the IceCube neutrino events and blazars (see \eg~\Refs~\cite{Krauss14,Padovani16,AhlersEx,Greek14,Petropoulou1603,Diltz16,Righi1607,Halzen16} and references erein), as well as studying diffuse neutrino emission \citep{Murase14,Dermer14}. 

Blazars significantly contribute to the diffuse (extragalactic) $\gamma$-ray background~\citep{2015ApJ...800L..27A}. If these $\gamma$-rays originate from proton interactions, the energy budget will be in principle sufficient to account for the intensity of IceCube neutrinos~\citep{Murase13}. However, recent stacking analyses using IceCube data suggest that blazars at most contribute 7-27\% of the observed neutrino intensity~\citep{Aarsten2017}. Similar arguments apply to other promising source candidates, such as radio galaxies~\citep{Hooper2016,Becker2014}, starburst galaxies~\citep{MuraseHidden16,Bechtol:2015uqb}, and Gamma-Ray Bursts~\citep{Abbasi:2012zw,Aartsen:2014aqy}.  While the origin of the astrophysical neutrinos is still unknown, the known constraints still permit that at least about ten percent of the observed neutrinos were produced by blazars~\citep{Aarsten2017}, possibly even by a few particularly neutrino-bright blazars.  

Blazars typically exhibit a two-hump structure in their spectral energy distribution (SED)~\citep{Fossati98,Ghi16}. This structure has been successfully reproduced by both leptonic and hadronic models for a number of blazars~\citep{Boettcher1304}. In a leptonic model, the low-energy and high-energy humps of the SED are produced by the synchrotron emission from electrons and the inverse-Compton scatter of soft photons (such as synchrotron photons from the same electrons), respectively. In a hadronic model, the primary electron-synchrotron generates the low-energy hump as well, but secondaries from hadronic processes induced by $p\gamma$ interactions are responsible for the $\gamma$-ray emission.  In the simplest cases, it is frequently used that in $p\gamma$ interactions the energy deposited in neutrinos (from $\pi^{\pm}$ decays) is roughly comparable to that of $\gamma$-rays (from $\pi^{0}$ decays), which means that the neutrino and $\gamma$-ray fluxes are directly correlated. Note, however, that even in hadronic models, the leading mechanism to generate the second hump may be, for instance, synchrotron radiation of secondary electrons produced by the photohadronic interactions. In addition, it may be possible that for certain astrophysical objects the emission in the second hump must be dominated by processes of purely leptonic origin. The study of this direct correlation is therefore the main motivation of this work.

One interesting test case for the direct neutrino--$\gamma$-ray correlation is the flat-spectrum radio quasar (FSRQ) PKS B1424-418. In \cite{Kadler16} (K16),  a positional and temporal coincidence between the 2-PeV neutrino event (IceCube event 35, IC35, or ``big bird'') and a burst of PKS B1424-418 was reported. After analyzing the $\gamma$-ray fluences from blazars in the positional-uncertainty region of IC35 as well as the diffuse $\gamma$-ray emission, it was  concluded that the burst of PKS B1424-418 had sufficient energy to account for the IceCube event, while the probability of a chance coincidence was around $5\%$. The required neutrino production efficiency was obtained by scaling the measured neutrino output to the $\gamma$-ray fluence in the energy range from $5~{\rm keV}$ to $10~{\rm GeV}$ ($10^{18.1}$--$10^{24.4}~{\rm Hz}$) and found consistent with that theoretically expected on the grounds of flavor, spectral effects, and source-population selection. An important ingredient was the assumption that the SED in that respective energy range were of hadronic origin, implying our direct correlation. We study if this assumption can be maintained in a self-consistent ansatz. Note that while we focus on one astrophysical object in this study, our conclusions will be more profound, as this direct correlation is widely used in multi-messenger analyses, see \eg~\cite{Turley:2016ver}.

Hadronic model of blazars typically involve many parameters \citep{Boettcher1304}, even in the simplest case: a parameter set for primary electrons, a set for protons, and a set for the bulk features of the emitting region, such as size (radius), magnetic-field strength, and Doppler factor, since none of these can be robustly derived from first principles or direct observations.  The characteristics for the SEDs vary greatly among the blazars \citep{Ghi16}, and therefore the best-fit parameters, even for purely leptonic models, exhibit large variations; an example are the leptonic models \cite{Finke08} versus \cite{Tavecchio13} for PKS B1424-418. Most neutrinos are produced in $p\gamma$ interactions near the $p\gamma$ threshold, and therefore the neutrino yield strongly depends on the density and spectrum of target photons. Even the simple assumption used in many studies, namely $L_\mathrm{\nu}$ $\propto$ $L_\mathrm{\gamma}$ (see \eg \cite{Righi1607,Aarsten2017}), is questionable, and an accurate and self-consistent calculation of the SED and the neutrino spectrum is necessary. The calculation also needs to be efficient to permit scanning a large parameter space. In this paper we present such a code and use it together with analytical calculations to explore what kind of model and which model parameters provide a consistent description of both the SED and the rate of PeV neutrinos.

The paper is structured as follows: in \Sec~\ref{ana} we introduce the general setup, and we use analytical calculations to determine the kind of model viable for PKS B1424-418; in \Sec~\ref{num} we briefly introduce our numerical methods and we present the results: the quality of the SED fit, the likelihood of having the observed neutrino event, and the corresponding parameters. The implications of the results are discussed in \Sec~\ref{conclusion}. The technical details on the analytical calculation, kinetic equations for the simulations and the numerical treatments are described in Appendix.\ref{appendix:ana},\ref{appendix:kin} and \ref{appendix:num}, respectively. We use cgs units in the paper, unless specified otherwise.

\section{General Analysis and Models}
\label{ana}

In this section we broadly list possible models and give generic constraints based on the characteristics of the SED for the source. The conclusions are derived using analytical and semi-analytical methods. 

\subsection{Assumptions and List of Models}

We use a one-zone model consisting of an isotropic, homogeneous and spherical emission region, or blob, with radius ${R_\mathrm{blob}^{\prime}}$, that moves relativistically with Doppler factor $\Gamma_\mathrm{bulk}$.  Electrons and protons are injected with power-law spectra, $d^{2}n^{\prime}/d\gamma^{\prime}dt^{\prime}=K^{\prime}\gamma^{\prime ~ \alpha}$ for $\gamma_\mathrm{min}^{\prime}<\gamma^{\prime}<\gamma_\mathrm{max}^{\prime}$, where $d^{2}n^{\prime}/d\gamma^{\prime}dt^{\prime}$ is the differential particle injection rate per volume, $\alpha$ the power-law index, $\gamma_\mathrm{min}^{\prime}$ and $\gamma_\mathrm{max}^{\prime}$ the minimum and maximum Lorentz factor of the particles, and $K^{\prime}$ is a normalization factor determined by the particle injection luminosity, $L_\mathrm{inj}^{\prime}$.  The injected electrons are henceforth referred to as primary electrons, whereas we denote as secondary electrons those created by hadronic interactions or $\gamma\gamma$ pair production. We allow primary electrons and protons to have separate parameter values for $L_\mathrm{inj}^{\prime}$, $ \gamma_\mathrm{min}^{\prime}$, $\gamma_\mathrm{max}^{\prime}$, and $\alpha$. The emission region is assumed to be filled with a homogeneous, randomly oriented magnetic field of strength $B^{\prime}$. Neutrinos and ``optically-thin'' photons can freely stream out of the blob on the timescale $t_\mathrm{fs}^{\prime}=3R^{\prime}/4c$. For simplicity, we assume the escape rate for charged particles, in the slow cooling case, to be a fixed multiple of the free-streaming timescale, $t_\mathrm{esc}=t_\mathrm{fs}/f_\mathrm{esc}=10 \, t_\mathrm{fs}$.
\footnote{This is clearly an oversimplification as the modeling of the particle escape requires a detailed specification of the geometry, the boundary conditions,  the magnetic-field configuration, \etc. The escape rate itself is likely to be energy-dependent on account of energy-dependent diffusion. In fact, synchrotron emission of escaped electrons may be responsible for the extended emission regions seen in VLBI radio data, and we do include such a component with the corresponding escape rate to fit the radio data in \Fig~\ref{fig:rates}.  Note that to some degree, the effect of energy-dependent escape can be compensated with an appropriate choice of the other free parameters such as $L_\mathrm{e,inj}^{\prime}$ and $\alpha_\mathrm{e,inj}$. See also \cite{Chen15,Chen16} for blazar models including an explicit treatment of diffusive escape and its effect on the particle density. Another possible scenario is to include an adiabatic cooling term which dominates over the escape rate, and this effect shows up effectively as an energy-independent extinction term in the kinematic equations. One implication is an explicit time dependence of particle densities arising from expansion. Besides, a steady-state can be reached only for specific geometries such as a stationary perturbation in an expanding flow, and  modeling the SED in this way is beyond the scope of the paper.}

The list of relevant interactions includes synchrotron emission and synchrotron self-absorption (SSA), inverse Compton (IC) scattering by both electrons and protons, $\gamma\gamma$ pair production and annihilation, Bethe-Heitler photo-pair production $p+\gamma \rightarrow p + e^{\pm}$ (BH), and photo-hadronic ($p\gamma$) interactions ($X+\gamma \rightarrow X^{\prime} + \pi$), where $X$ and $X^{\prime}$ denotes either a proton or a neutron and $\pi$ includes charged or neutral pions. Secondary particles such as $\pi^{\pm}$ and $\mu^{\pm}$ can in principle radiate before they decay, but for the parameter values relevant to this study the effect is negligible. The details are listed in \Tab\ref{AppendixTable} and described in Appendix.\ref{appendix:kin}.

The low-frequency hump in the SED of a generic blazar extends from the radio band to the UV and in some cases even to the X-ray band; the high-frequency component can be observed from X-rays up to TeV $\gamma$-rays. Here we discuss four scenarios that may in principle account for shape of the SED: the pure leptonic (SSC) model, the lepto-hadronic Synchrotron-Self-Compton (LH-SSC) model, the lepto-hadronic pion (LH$\pi$) model, the lepto-hadronic proton-synchrotron model and the proton model, which is a purely hadronic model. The defining features of these models are summarized in \Tab~\ref{modellist}.


\begin{table*}[h]
	\centering
	\begin{tabular}{|c|c|c|c|}
		\hline
 		& First peak	& Middle range		& Second peak \\
 		& (eV-keV)		& (keV-MeV) 		& (MeV-TeV) \\
		\hline
		{\bf SSC} & {\bf L} & {\bf L} & {\bf L} \\
		(Pure leptonic) & Primary $e^{-}$ synchrotron & SSC & SSC \\
		\hline
		{\bf LH-SSC} & {\bf L} & {\bf H} & {\bf L} \\
		(Lepto-hadronic) & {Primary $e^{-}$ synchrotron} & Secondary leptonic  & SSC by primary $e^{-}$\\
		\hline
		{\bf LH-$\boldsymbol{\pi}$} & {\bf L} &{\bf H} & {\bf H} \\
		(Lepto-hadronic) & {Primary $e^{-}$ synchrotron} & Secondary leptonic  & \parbox[t]{4cm}{Secondary leptonic or $\gamma$-rays  from direct $\pi^{0}$ decay}\\
		\hline
		{\bf LH-psyn}  & {\bf L} & {\bf H} & {\bf H}\\
		(Proton synchrotron)  & {Primary $e^{-}$ synchrotron} & \parbox[t]{3cm}{Proton synchrotron or secondary leptonic} & Proton synchrotron \\
		\hline
		{\bf Proton}  & {\bf H} & {\bf H} & {\bf H} \\
		(Pure hadronic) & Proton synchrotron & Secondary leptonic & \parbox[t]{4cm}{Secondary leptonic or $\gamma$-rays from direct $\pi^{0}$ decay}\\
		\hline
	\end{tabular}
	\caption{List of models. In the table, ``{\bf L}''=``Leptonic'' and ``{\bf H}''=``Hadronic''. ``LH'' is the abbreviation for ``lepto-hadronic'', which is a mixture of leptonic and hadronic components.}
	\label{modellist}
\end{table*}


In both the SSC and LH-SSC model, the first hump is described by synchrotron emission of primary electrons, and the high-frequency component is due to inverse Compton scattering of those photons by the same electrons.  The LH-SSC model contains an additional hadronic component compared to the pure leptonic SSC model, which may fill the gap between the two humps (\eg, accounting for the X-ray emission from PKS B1424-418). 

In the ``Lepto-hadronic'' models internally, the low-frequency hump is likewise described by synchrotron emission of primary electrons, whereas the second hump arises from hadronic processes. Depending on the parameters of the source, the dominant contribution to the second peak can be $\gamma$-rays from $\pi^{0}$ decays, synchrotron and inverse Compton radiation emitted by $e^{\pm}$ from $\pi^{\pm}$ decays, internal $\gamma\gamma$ annihilation, or proton-pair production (Bethe-Heitler, BH), where it can be called an ``$\mathrm{LH-\pi}$'' model. The second peak can also be produced in some cases by the proton-synchrotron model (e.g. FSRQ 3C 279 \citep{Diltz15}), namely the ``LH-psyn'' model.

In the proton model, leptonic emission is sub-dominant at all wavebands. The first peak in the SED is attributed to proton-synchrotron radiation, and the second peak is produced by the same type of hadronic processes in the LH$\pi$ model.

\subsection {Constraints on Models and Parameters from Semi-analytical Calculations}
\label{modelconstraints}

While both the leptonic and the hadronic models have successfully explained the SED of a number of blazars, for example Mrk 421, 3C 279 etc., the unique combination of the SED with the PeV-neutrino information places stringent limits on the model of PKS B1424-418. Here we use analytical and semi-analytical calculations to demonstrate that neither the lepto-hadronic (LH$\pi$, LH-psyn) nor the purely hadronic proton model can simultaneously explain the SED and the PeV-neutrino event, leaving the SSC and LH-SSC models of the SED as the only viable contenders. Analytical arguments also give useful constraints on the parameter space.


{\bf Proton model}:
In p$\gamma$ interactions, the neutrino energy is roughly 5\% of that of the parent proton, $E_\mathrm{\nu} \sim 0.05 \, E_{p}$. For PKS B1424-418 at the redshift $z=1.522$, the Lorentz factor of the proton in the comoving frame can be written in terms of the bulk Lorentz factor of the blob $\Gamma$  and the neutrino energy in the observer frame $E_\mathrm{\nu,PeV}^\mathrm{ob}\equiv E_\mathrm{\nu}^\mathrm{ob}/{\rm PeV}$ as 
\begin{equation}
	\label{equ:gamma_p}
	\gamma_\mathrm{p}^{\prime}\sim5\times10^{7}\,\Gamma^{-1}\,{\rm E}_\mathrm{\nu,PeV}^\mathrm{ob}\ .
\end{equation} 
If the low-energy peak is attributed to proton synchrotron emission, the peak frequency of the synchrotron emission, $\nu_\mathrm{pk,1}$, obeys 
\begin{equation}
	\label{equ:peak_freq}
	h\nu_\mathrm{pk,1}^{\prime}=m_\mathrm{e}c^{2}\frac{m_\mathrm{e}}{m_\mathrm{p}}\left(\frac{B^{\prime}}{B_\mathrm{crit}}\right)\,{\gamma_\mathrm{p}^\prime}^{2}\,
\end{equation} 
where $B_\mathrm{crit}\equiv4.41\times10^{13}$~G is the critical magnetic field. Combining \eqs~(\ref{equ:gamma_p}) and (\ref{equ:peak_freq}) provides a constraint on the magnetic field, 
\begin{equation}
	\label{equ:synpeak}
	B^{\prime}|_\mathrm{pk,1}=(7\times10^{-4}\ \mathrm{G})\, \left(\frac{\nu_\mathrm{pk,1}^\mathrm{ob}}{\rm 10^{14}\ Hz} \right) \left(\frac{\Gamma_\mathrm{bulk}}{10} \right) \left(\frac{{E}_\mathrm{\nu}^\mathrm{ob}}{\rm PeV} \right)^{-2}\ .
\end{equation} 
The peak frequency of the second hump in the SED relates to the peak energy of the secondary ${e^{\pm}}$ that result from either $\pi^{\pm}\rightarrow e^{\pm}$ decay or from $\pi^{0}\rightarrow \gamma\gamma \rightarrow e^{\pm}$ reactions. It will be shown later that for PKS B1424-418, up to one generation of $e^{\pm}$ cascade is expected, so that for both channels the energy of $e^{\pm}$ is  $E_\mathrm{e}\sim 0.05 E_\mathrm{p}$. Reproducing the second hump with the observed peak frequency $\nu_\mathrm{pk,2}^\mathrm{ob}$ from synchrotron emission of the secondary pairs requires a magnetic-field strength 
\begin{equation}
	B^{\prime}|_\mathrm{pk,2}=(3\times10^{-2}\ \mathrm{G})\, \left(\frac{\nu_\mathrm{pk,2}^\mathrm{ob}}{\rm 10^{23}\ Hz} \right) \left(\frac{\Gamma_\mathrm{bulk}}{10} \right) \left(\frac{E_\mathrm{\nu}^\mathrm{ob}}{\rm PeV} \right)^{-2}\ .
\end{equation} 
The value $B|_\mathrm{pk,2}$ is clearly incompatible with $B|_\mathrm{pk,1}$, which means that the proton model is not viable.


{\bf LH-psyn model}:
If we require a proton-synchrotron origin of the high-energy hump in the SED and a neutrino emission peaked at PeV-energies, the requirement on the magnetic field strength can be derived in a similar way as \equ{synpeak}:
\begin{equation}
	B_\mathrm{psyn}^{\prime}=(7\times10^{5}\mathrm{G})\left(\dfrac{\nu_{\mathrm{pk},2}^{\mathrm{ob}}}{10^{23} \mathrm{~Hz}}\right)\left(\dfrac{\mathrm{\Gamma_\mathrm{bulk}}}{10}\right)\left(\dfrac{\mathrm{E_{\nu}^{ob}}}{\mathrm{PeV}}\right)\ .
\end{equation}
The magnetic energy density is then of the order of $10^{8}\mathrm{~erg/cm^{3}}$. Compared to the photon energy density for this source $u_\mathrm{phot}\sim6\times10^{-6}L_{46}^{iso}R_{18}^{-2}\Gamma_{1}^{-4}\mathrm{~erg/cm^{3}}$, it is clearly unphysical for the jet energy budget.


{\bf LH-$\boldsymbol{\pi}$ model}: 
	This scenario is slightly more complicated, but a few generic conditions must be met: 
\begin{enumerate} 
	\item The proton synchrotron flux must not exceed that of synchrotron radiation of primary electrons.  \label{constraint:1}
	\item The $Y_\mathrm{SSC}$ parameter, defined as the power ratio of synchrotron-self Compton emission to synchrotron emission, must not exceed unity for this source (otherwise the model becomes the SSC model). \label{constraint:2}
	\item The observed peak frequency of the high-energy hump in the SED $\nu_\mathrm{pk,2}^\mathrm{ob}\sim10^{23}$~Hz must be consistent with the characteristic energy of the secondary $e^{\pm}$ from p$\gamma$ interactions $E_\mathrm{e}\sim 0.05 \, E_\mathrm{p}$, with $E_\mathrm{p}$ determined from the neutrino energy $E_\mathrm{\nu,PeV}^\mathrm{ob}$. \label{constraint:3}
	\item The emission from pairs by the Bethe-Heitler process must not overshoot the observation. \label{constraint:4}
\end{enumerate} 


\begin{figure*}[h]
	\centering 
	\includegraphics[width=0.75\columnwidth]{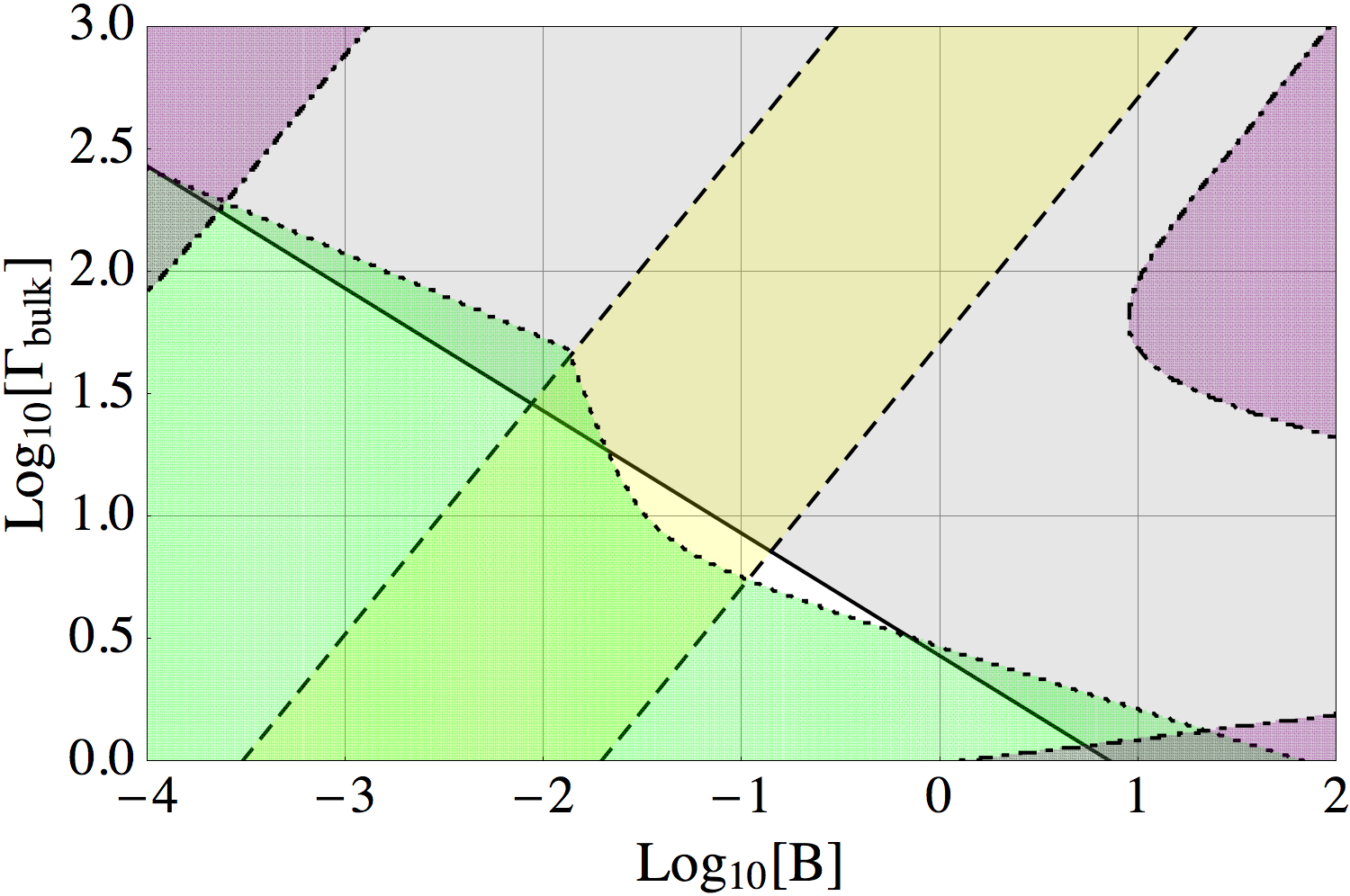}
	\caption{Allowed parameter regions for the LH$\pi$ model (\cf,\Sec~\ref{modelconstraints}). Green: lower-left region with dotted boundary, correponding to constraint \ref{constraint:1}; Grey: upper-right region limited by the solid line, constraint \ref{constraint:2}; Yellow area between the dashed parallel lines, constraint \ref{constraint:3}; Purple: the three separate regions formed by dot-dashed boundaries, constraint \ref{constraint:4}.}
	\label{lhpimodel}
\end{figure*}


Here the SED is approximated by four segments of power-law spectra. The equations to calculate these four constraints are \ref{constraint:1}:\equ{fpsyn}; \ref{constraint:2}:\equ{ysscestimate}; \ref{constraint:3}:\equ{pgammapeak} and \ref{constraint:4}:\equ{nufnubh} from Appendix.\ref{appendix:ana}. All four constraints are displayed in \Fig~\ref{lhpimodel} for a blob radius $R_\mathrm{blob}^{\prime}=10^{18}$~cm. When $R_\mathrm{blob}^{\prime}$ increases, the boundaries of the grey and green regions (corresponding to constraints 1 and 2, respectively) move towards the lower-left corner of the panel, whereas the regions defined by constraints 3 and 4 are independent of the $R_\mathrm{blob}^{\prime}$. There is no region of overlap for all four constraints, whatever the value of $R_\mathrm{blob}^{\prime}$, which rules out the LH-$\pi$ model for PKS B1424-418.



{\bf SSC and LH-SSC models}:
In the Thomson scattering regime, the frequency of the scattered photon is $\nu_\mathrm{pk,2}^{\prime}\simeq \gamma_\mathrm{e}^{\prime 2}\nu_\mathrm{pk,1}^{\prime}$. Scattering proceeds in the Klein-Nishina regime for electron Lorentz factors $\gamma^\prime\gtrsim \gamma_\mathrm{KN}^{\prime}$, here $\gamma_\mathrm{KN}^{\prime} h \nu_\mathrm{pk,1}^{\prime}\sim m_\mathrm{e}c^{2}$.  Combining the two expressions, we find the following relationship among the Lorentz factor of the accelerated electrons in the comoving frame $\gamma_\mathrm{e}^{\prime}$ and the SED parameters of PKS B1424-418: 
\begin{equation}
	\gamma_\mathrm{e}^{\prime}=0.007\, \gamma_\mathrm{KN}^{\prime}\,\left(\frac{\nu_\mathrm{pk,1}^\mathrm{ob}}{\rm 10^{14}\ Hz}\right)\left(\frac{\nu_\mathrm{pk,2}^\mathrm{ob}}{\rm 10^{23}\ Hz} \right) \left(\frac{\Gamma}{10} \right) ^{-1} < \gamma_\mathrm{KN}^{\prime}\ .
	\label{equ:KN}
\end{equation}
We conclude that the inverse Compton scattering of primary electrons is always Thomson scattering. 

The comoving photon density can be written as 
\begin{equation}
	u_\mathrm{ph}^{\prime}=\left(\frac{d}{R_\mathrm{blob}}\right)^{2} \left(\frac{1+z}{\Gamma^{2}}\right)^{2} \frac{\nu F_\mathrm{\nu}}{c} \ .
\end{equation} 
Here $d=4.477$~Gpc is the comoving radial distance of the source (For $z=1.522$ and a flat $\Lambda$CDM universe with $\Lambda=0.7$), and $u_\mathrm{B}^{\prime}=B^{\prime 2}/8\pi$ is the comoving magnetic-field energy density. We can express the constraints on $B^{\prime}$, $R_\mathrm{blob}^{\prime}$, and $\Gamma$ in terms of observed quantities as
\begin{equation} 
	\label{equ:SSCconstraint}
		R_\mathrm{blob}^{\prime}\Gamma  \approx (1.4\times10^{19}\	\mathrm{cm})\,Y_\mathrm{SSC}^{-1/2}\, \left( \frac{\nu F_\mathrm{\nu}|_\mathrm{pk,1}^\mathrm{ob}}{\rm 10^{-11}\ erg\, cm^{-2}\,s^{-1}} \right)^{1/2} \left( \frac{\nu_\mathrm{pk,1}^\mathrm{ob}}{\rm 10^{14}\ Hz}	\right)^{-2} \left( \frac{\nu_\mathrm{pk,2}^\mathrm{ob}}{\rm 10^{23}\ Hz}\right)  \,
\end{equation} and 
\begin{equation} 
	\label{equ:SSCconstraintB}
	B^{\prime}\approx (9.2\times10^{-3}\ \mathrm{G})\, \left(\frac{\nu_\mathrm{pk,1}^\mathrm{ob}}{\rm 10^{14}\ Hz} \right)^{2} \left(\frac{\nu_\mathrm{pk,2}^\mathrm{ob}}{\rm 10^{23}\ Hz} \right)^{-1} \left(\frac{\Gamma}{10} \right)^{-1} \, 
\end{equation} where
\begin{equation}
	Y_\mathrm{SSC}\approx \frac{\nu^{\prime} u_\mathrm{\nu}^{\prime}(\nu_\mathrm{pk,1}^{\prime})}{u_\mathrm{B}^{\prime}} \approx \frac{\rm \nu F_\mathrm{\nu}|_\mathrm{pk,2}} {\rm \nu F_\mathrm{\nu}|_\mathrm{pk,1}}\approx 10\ 
\end{equation} 
and $u_\mathrm{\nu}^{\prime}$ is the differential energy density of photons in the comoving frame.  The scaling values of the observed quantities are typical for the SED of PKS B1424-418 (\cf, Fig.\ref{fig:rates}). Therefore \equ{SSCconstraint} implies that the blob radius be large, on the order of a light-year. The magnetic-field strength must be rather low, suggesting that the blob is located far away from the central engine, beyond the broad-line region and the dusty torus (see also \cite{Tavecchio13}). This scenario justifies that we neglect inverse Compton scattering of external photons.


The optical depth of MeV-band $\gamma$-rays to pair production, $\tau_{\gamma\gamma}$, can be estimated as 
\begin{equation} 
	\tau_{\gamma\gamma}\approx \frac{\rm L^{\prime}}{ R_\mathrm{blob}^\prime} \, \frac{\sigma_{T}}{4\pi m_{e}c^{3}} \approx 3\times10^{-5}\, \left( \frac{\nu F_{\nu}^{ob}}{\rm 10^{-11}\ erg\, cm^{-2}\, s^{-1}} \right) \left( \frac{\Gamma}{10} \right)^{-4} \left( \frac{R_\mathrm{blob}^\prime}{\rm 10^{18}\ cm} \right)^{-1}\ ,
\end{equation} 
indicating that MeV $\gamma$-rays can escape the source. A numerical calculation on the optical depth shows that $\tau_{\gamma\gamma}$ approaches unity for multi-TeV to PeV $\gamma$-rays for typical model parameters for PKS B1424-418, which is consistent with one cascade generation of these $\gamma$-rays.

The position and flux of the high-energy peak in the SED need to be explained by the SSC model. Determining the permitted abundance of hadrons as well as the detailed fitting of the entire SED and the neutrino data require the numerical modeling of the source, which is discussed in the following section.

PKS B1424-418 exhibits significant time variabilities over a wide range of scales, as other FSRQs and BL Lacs do. Variability timescales $t_\mathrm{var}^\mathrm{ob}$ as short as a month are consistent with the causality argument $t_\mathrm{var}^\mathrm{ob}\sim R_\mathrm{blob}^{\prime}(1+z)/\Gamma c$ and the constraint set by
\equ{SSCconstraint} for a blob size $R_\mathrm{blob}=7.5\times10^{17}$~cm and a bulk Lorentz factor $\Gamma=35$, for which $t_\mathrm{var}^\mathrm{ob}\simeq 3$~weeks.\footnote{This choice of $\Gamma$ is also in line with the best-fit spectra by numerical simulations over the parameter space.} Faster variability has been observed on timescales shorter than a day, which may be explained by compact substructures in the jet, such as re-collimation \citep{Bromberg09}, ``jet in a jet'' scenarios, or magnetic reconnection \citep{Giannios09,Giannios13}. In any case, explaining this very fast variability is beyond the scope of this paper. 


\section{Numerical SED Model and Consequences for the Neutrino Production}
\label{num}

We present our numerical simulation results for the  SSC and LH-SSC scenarios here, which were found to be preferable in the previous section. After discussing our methods, we will first discuss the Burst phase, and then show a self-consistent picture for the evolution of the blazar over the full studied timeline.

\subsection{Numerical Methods}


\begin{figure*}
	\centering 
	\includegraphics[width=0.75\columnwidth]{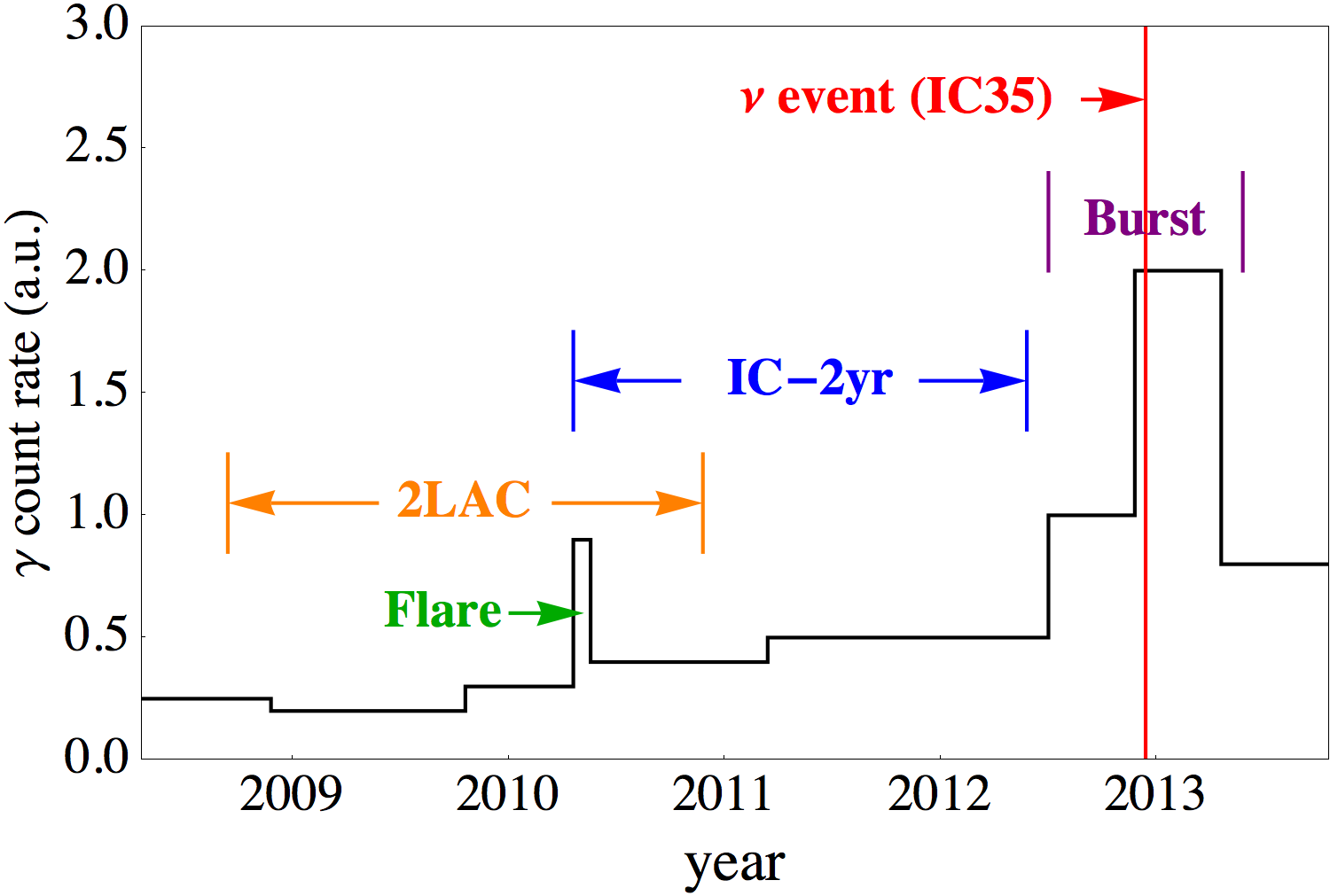}
	\caption{An illustration of the $\gamma$-ray count rate as a function of time together with a definition of the 2LAC, IC-2yr, and Burst phases of PKS B1424-418. Note that the count rate is meant for illustrative purposes only and does not accurately reflect the data (see Figure 1 of K16 for the bi-weekly binned $\gamma$-ray light curve). Also note that K16 shows the SED from the IceCube 3-year phase, which overlaps with the Burst phase. The IC-2yr SED is constructed and provided by the authors of K16, which is not shown in their paper.} 
	\label{illus} 
\end{figure*}



\begin{table*}[tbp]
	\centering
	\begin{tabular}{|c|c|c|p{10cm}|}
		\hline
 		~& Group & Symbol & Definition \\
		\hline
 		~&~& $R_\mathrm{blob}^{\prime}$ & Comoving radius of blob, fixed to $7.5\times10^{17}$ cm\\
 		~& Global & $f_\mathrm{esc}$ & $e^{\pm}$ and $p$ escape fraction, fixed to $1/10$ \\
 		~&~& $\Gamma_\mathrm{bulk}$ & bulk Lorentz factor of the blob fixed to $35$\\
		\cline{2-4}
 		~&~& $B^{\prime}$ & Magnetic field strength, blob frame \\
 		~&~& $L_\mathrm{e,inj}$ & Injection luminosity of primary $e^{-}$, AGN frame\\
 		Parameters & Leptonic & $\gamma_\mathrm{e,min}^{\prime}$ & Minimum Lorentz factor of primary $e^{-}$, blob 	frame\\
 		~&~& $\gamma_\mathrm{e,max}^{\prime}$ & Maximum Lorentz factor of primary $e^{-}$, blob frame\\
 		~&~& $\alpha_\mathrm{e,idx}^{\prime}$ & Power law index of injected primary $e^{-}$\\
		\cline{2-4}
 		~&~& $\alpha_\mathrm{p,idx}^{\prime}$ & Power law index of injected protons, fixed to $-2.0$\\
 		~& Hadronic & $\eta_\mathrm{b}$ & Luminosity ratio $p$ to $e^{-}$ at injection, \ie, $\eta_\mathrm{b}\equiv L_\mathrm{p,inj}/L_\mathrm{e,inj}$ \\
 		~&~& $E_{p,max}^{ob}$ & Maximal energy of injected protons, observer frame\\
		\hline
					~&\multicolumn{2}{c|}{$P_{0,1,0}(E_\mathrm{p,max},\eta_\mathrm{b})$}& Probability to observe 0,1,0 neutrino 
					events in the 0.5-1.6~PeV, 1.6-2.4~PeV and $>2.4$~PeV bands in IceCube, respectively,
					as a function of $E_\mathrm{p,max}$ and $\eta_\mathrm{b}$\\
					~&\multicolumn{2}{c|}{$P_\mathrm{\nu,max}$}&{Maximum value of $P_{0,1,0}(E_\mathrm{p,max},\eta_\mathrm{b})$ in the
					parameter space}, named as ``neutrino best-fit''\\
					~&\multicolumn{2}{c|}{$L_\mathrm{\nu}$}&{Total neutrino luminosity, 
					including all flavors}\\
					~&\multicolumn{2}{c|}{$L_\mathrm{\gamma}$}&{$\gamma$-ray luminosity integrated over the frequency band $10^{18.1}\sim10^{24.4}~$Hz}\\
		Notations	~&\multicolumn{2}{c|}{$+$}&{SED best-fit mark}\\
					~&\multicolumn{2}{c|}{$\times$}&{Neutrino best-fit mark, global}\\
					~&\multicolumn{2}{c|}{$\otimes$}&{Neutrino best-fit under the constraint of
					SED being reproduced within $3\sigma$ confidence}\\
					~&\multicolumn{2}{c|}{\ding{83}}&{Joint best-fit mark, for SED and neutrino}\\
					~&\multicolumn{2}{c|}{\ding{172}}&{Neutrino flux to expect one $>0.5~$PeV 
					event during 2LAC phase, assuming the same effective area of IceCube as in 
					IC-2yr and Burst phase (see \Fig~\ref{fig:2lac}}) \\
		\hline					
	\end{tabular}
	\caption{List of parameters and notations.}
	\label{paratable}
\end{table*}


We simulate time-dependent particle spectra for $e^{\pm}$, $p$, $n$, $\gamma$, and $\nu_{\alpha}$ ($\alpha$ denotes the neutrino flavor) by numerically solving the time-dependent differential-integral kinematic-equation system in the energy space $\gamma$, for all the particle species mentioned above:

\begin{equation} 
	\partial_{t}n(\gamma,t)=-\partial_{\gamma}\{{\dot{\gamma}(\gamma,t)n(\gamma,t)-\partial_{\gamma}[D(\gamma,t) n(\gamma,t)]/2}\}-\alpha(\gamma,t) n(\gamma,t)+Q(\gamma,t)
	\label{eqn:kinmain}
\end{equation}

where $n(\gamma)\equiv d^{2}N/d\gamma dV$ is the differential number density of the particle species. In the equation above, the source term $Q$ may depend on energy $\gamma$, time $t$, and the current target-particle distributions $\{n_{tar}(\gamma)\}$. It models the injection, emission or generation of a new particle after an interaction, or re-distribution or re-injection of the same particle after scattering. The sink term $\alpha$ depends on same set of variables and functionals, which models the escape of the particle from the blob, decay, annihilation and disappearance due to an interaction. 
The differential terms $\dot{\gamma}(\gamma,t)$ and $D(\gamma,t)$ account for the particle cooling and diffusion effect in the momentum space due to synchrotron radiation, Thomson scattering and Bethe-Heitler process. In those processes, electron or proton loses a tiny fraction of energy after scattering, and the finiteness of the numerical grid spacing is unable resolve this tiny shift in the redistribution function, via the terms $\alpha$ and $Q$. Therefore, we apply the ``continuous-loss'' approximation, demanding an accuracy up to the second-order differentiation. Due to isotropy and spherical symmetry, only the radial component $D(\gamma,t)$ of the diffusion tensor appears in the equation. 
\footnote{The mathematical forms of this treatment are expressed as \eqs~(\ref{equ:continuousloss}-\ref{equ:dic}), \eqs~(\ref{equ:gammadotsyn}-\ref{equ:Hsyn}) and \equ{dbh}. }

The rates and redistribution functions are described by physics, where we consider synchrotron, inverse Compton, pair production and annihilation, photo-hadronic ($p\gamma$) interaction and Bethe-Heitler (photo-pair) process. See Appendix.\ref{appendix:kin} for details and \cite{HU10} (H10) for the simplified $p\gamma$ interaction model which we apply in this paper.

Bremsstrahlung and $pp$ collisions are neglected in our case, which can be relevant in blazars when the emission region is compact \citep{Eichmann2012}.
However, in our case, the large blob-size leads to a low number density of cold protons, which can be easily estimated as $n_\mathrm{p,cold}\sim2.5\times10^{-5}\eta_\mathrm{p,cold}R_{18}^{-2}\Gamma_{1.5}^{-4}L_{46}^{ob}$. Here $L^{ob}$ is the photon luminosity in the observer frame and we have parameterized the energy density of cold protons as a multiple $\eta_\mathrm{p,cold}$ to that of photons. Adopting the zeroth-order approximation on the $pp$ cross-section, $\sigma_\mathrm{pp}\sim50 \mathrm{~mb}$, the optical depth of $pp$ collision is estimated to be $\tau_{pp}\sim 10^{-12}\eta_\mathrm{p,cold}R_{18}^{-1}\Gamma_{1.5}^{-4}L_{46}^{ob}$, which is negligible. The energy-loss rate due to Bremsstrahlung is estimated by \citep{Eichmann2012} as $\dot{\gamma}_\mathrm{brem}/\gamma\sim1.4\times10^{-16}(\ln2-1/3)n_\mathrm{p,cold}\sim1.3\times10^{-21}\eta_\mathrm{p,cold}R_{18}^{-2}\Gamma_{1.5}^{-4}L_{46}^{ob}$, which is significantly below that of synchrotron loss-rate $\dot{\gamma}_\mathrm{syn}/\gamma\sim10^{-12}B_{-3}^{2}\gamma_{3}$ even for the lowest-energy electrons in our case.

We use the finite-difference method to solve the equation numerically, on an evenly-spaced logarithmic grid in energy and a linear one in time. The ``Crank-Nicolson'' differential scheme is used in time with the ``Chang \& Cooper'' \citep{Chang1970} scheme in energy, to achive stability and a more accurate goal in the correct steady-state solution. For the compatibility with the latter scheme as well as increased accuracy, we calculate up to the second-order differential term from physics. See Appendix.\ref{appendix:num} for details and \cite{VP08} (VP09) for the application to leptonic processes. Our list of input parameters and assumptions is summarized in \Tab~\ref{paratable}, which we describe in greater detail in this and the next sections.

The dynamical SEDs of PKS B1424-418 reported in K16 are categorized into four phases: (1) flare in 2010, lasting about 1 month; (2) 2LAC phase, from 2008.8 to 2010.9; (3) IC-2yr period, from 2010.5 to 2012.5, the first two years of IceCube observation; (4) Burst phase, from 2012.6 to 2013.3, when the source experienced a long-lasting high-flux phase in $\gamma$-rays. \Fig~\ref{illus} provides a visual timeline including the time-averaged GeV-band $\gamma$-ray flux. A 2-PeV neutrino event in IceCube (IC35, also dubbed ``big bird'') was observed on Dec. 4, 2012, during phase 4, with a position consistent with that of PKS B1424-418.  The SEDs shown in K16 are based on time-averaged spectra from each phase.  In this paper, the flare phase is ignored since its duration was too short to result in a neutrino fluence comparable with that of other phases.\footnote{Even for the most optimistic estimates in K16, the expected neutrino count is far below 1.0, consistent with the null detection of neutrinos during this phase.}  We therefore focus on three phases (2LAC, IC-2yr, and Burst), as indicated in \Fig~\ref{illus}, which can be interpreted as the time-dependent evolution of the AGN. We will first simulate the phases independently, and then interpret the evolution (changes) of the parameters.

For each phase, the SED and neutrino spectra are modeled by a steady-state solution to \eq\ref{eqn:kinmain}. Under the SSC and LH-SSC scenarios, the first peak of the SED is described as synchrotron emission from primary $e^{-}$, and the $\gamma$-rays are described as SSC emission from the same $e^{-}$ population (see \Tab~\ref{modellist}).  After fixing a few global parameters such as $R_\mathrm{blob}$ and $\Gamma_\mathrm{bulk}$, the following two-step simulations are performed for each of the three phases of PKS B1424-418 (2LAC, IC-2yr, and Burst): 1) Use leptonic simulations $(\eta_b \equiv 0)$ to find the best-fit parameters of primary $e^{-}$ for the low-energy hump and $\gamma$-ray band ($10^{22}-10^{25}$ Hz). 2) Inject protons, until their spectrum reaches a steady state to find the total SED and neutrino spectrum.

In step 1, with leptonic simulations, the following parameter space is scanned: $L_\mathrm{e,inj}(10^{42.5}\sim10^{45.5}\mathrm{~erg/s}) \otimes \gamma_\mathrm{e,min}^{\prime}(10^{2.6}\sim10^{3.9}) \otimes \gamma_\mathrm{e,max}^{\prime}(10^{4.2}\sim10^{6.0}) \otimes \alpha_\mathrm{e,inj}(-2.0\sim-1.0) \otimes B^{\prime}(10^{-2.2}\sim10^{-4.0}\mathrm{~G}) \otimes \Gamma_\mathrm{bulk}(1\sim200)$. The best-fit parameters are obtained by $\chi^{2}$-minimization.

In step 2, including protons, the SED and the neutrino spectrum are calculated for a logarithmic $\eta_\mathrm{b} (10^{2}\sim10^{8})\otimes E_\mathrm{p,max}^{\prime}(10^{2}\sim10^{9}\mathrm{~GeV})$ parameter grid for each phase of PKS~B1424-418.  Here $\eta_\mathrm{b}$ is the proton-to-electron luminosity ratio (baryonic loading) at injection, $\eta_\mathrm{b}=L_\mathrm{p,inj}/L_\mathrm{e,inj}$, and $E_\mathrm{p,max}^{\prime}$ is the maximum proton energy. Each point of the grid corresponds to a unique combination of $\eta_\mathrm{b}$ and $E_\mathrm{p,max}$ and therefore an independent simulation. The entire grid has a resolution of $400\times160$ for each phase, requiring a total of 192000 hadronic simulations.

The global parameters $R_\mathrm{blob}^\prime=7.5\times10^{17}$ cm and $\Gamma_\mathrm{bulk}=35$ are chosen according to \equ{SSCconstraint}, including a variability time $t_\mathrm{var}<1$~month.  The energy-independent escape rate for $e^{-}$ and $p$ is assumed to be $t_\mathrm{e,p}^{-1}=0.1 \, t_{\text{fs}}^{-1}$, which is an order magnitude slower than the free-streaming escape rate (that of the neutrinos).  We also use a fixed power law injection index $\alpha_\mathrm{p,inj}=-2.0$ for proton injection, where the results are not very sensitive to this value.

TANAMI VLBI data of PKS B1424-418 indicate an angular size of a few milli-arcsecond for the emission region in the radio band, which translates into a physical size of about the order of a few $10^{19}$~cm -- which is larger than that of the blob. The extended VLBI component can be interpreted as synchrotron emission from electrons that escaped from the blob into an extended region of size $R_\mathrm{ext} \sim 3.0\times10^{19}$~cm that is filled with weaker magnetic field $B_\mathrm{ext}^{\prime}\simeq 0.1 \, B_\mathrm{blob}^{\prime}$. We assume that the radio data are fitted by synchrotron emission from electrons that escaped the blob into an extended region using these parameters. All other contributions to the SED come from the blob.


\begin{figure*}
	\centering
	\includegraphics[width=1.0\columnwidth]{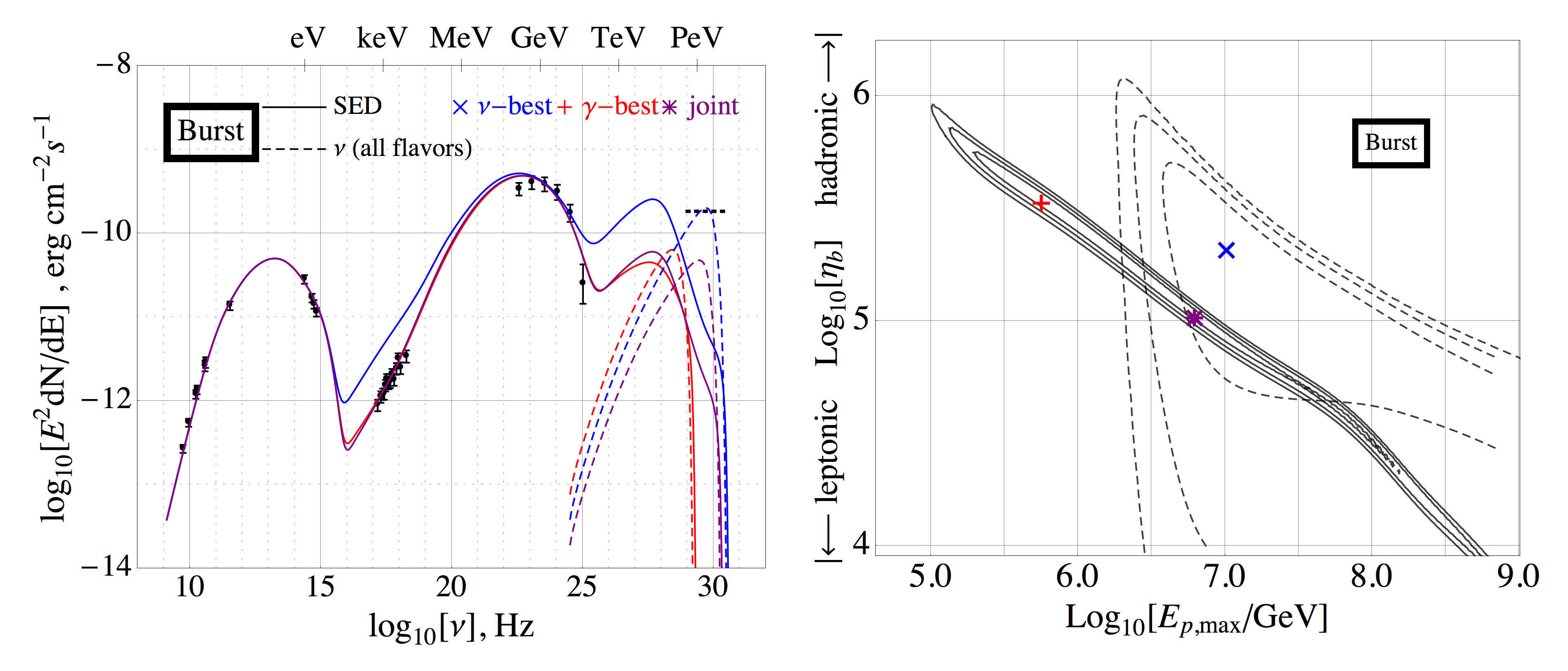}
	\caption{Right panel: fitting quality to SED and neutrino observations over the $E_{p,max}$ and $\eta_{b}=L_{p,inj}/L_{e,inj}$ parameter-space for the Burst phase.  Solid contours are the boundaries of $1\sigma$, $2\sigma$, and $3\sigma$ confidence regions for the SED fit, while dashed curves are iso-probability contours for $P/P_\mathrm{\nu,max}=0.32, 0.05, 0.003$ -- see main text for details.  The symbols $+$, $\times$ and \ding{83} mark the best-fit parameters of the SED, neutrino and the joint fit, respectively. Left panel: the SED (solid) and neutrino (dashed) spectra, corresponding to the three marks on the right-panel. The horizontal dashed line is the neutrino flux level to observe one PeV event over this 9-month period. Data points are provided from authors of K16.}
	\label{fig:burst}
\end{figure*}


\subsection{SED Model and Neutrino Production, the Burst Phase}
\label{sec:burst}

In this subsection, we focus on the Burst phase and the relationship to the potentially observed neutrino event during that phase.

For each parameter-space simulation, we perform independent optimizations. First, we adapt our model to the SED and find the best fit and confidence regions by calculating the reduced $\chi^{2}$-values
\footnote{The data points and statistical error-bars are provided by authors of K16. For total errors, we supplement an estimated systematic error of around 10\% for radio to X-ray bands so that the data can be well described by power-laws. For Fermi-LAT, the systematic error is dominated by uncertainties of the effective area for {\it Pass7} data, which is rouphly 10\% to 15\%, depending on energy.}.
This SED best-fit is marked with the symbol ``$+$''. Then with the predicted neutrino spectrum from each simulation, we calculate the probability, $P_{010}(E_\mathrm{p,max},\eta_\mathrm{b})$ to observe from PKS B1424-418 with IceCube 0,1, and 0 neutrino events, as indeed observed, in the $(0.5-1.6),(1.6-2.4),$ and $>2.4$ PeV energy bands, respectively. In each band, the expected number of neutrino events follows a Poisson distribution. The best adaption to the neutrino data without regard of the SED is referred to by the symbol ``$\times$'', and its fit probability is denoted as
$P_\mathrm{\nu,max}$. The joint best-fit point, by maximizing the joint probability of the SED and neutrino fit, is marked with the symbol ``\ding{83}''. 


\begin{figure*}[h]
	\centering
	\includegraphics[width=0.75\columnwidth]{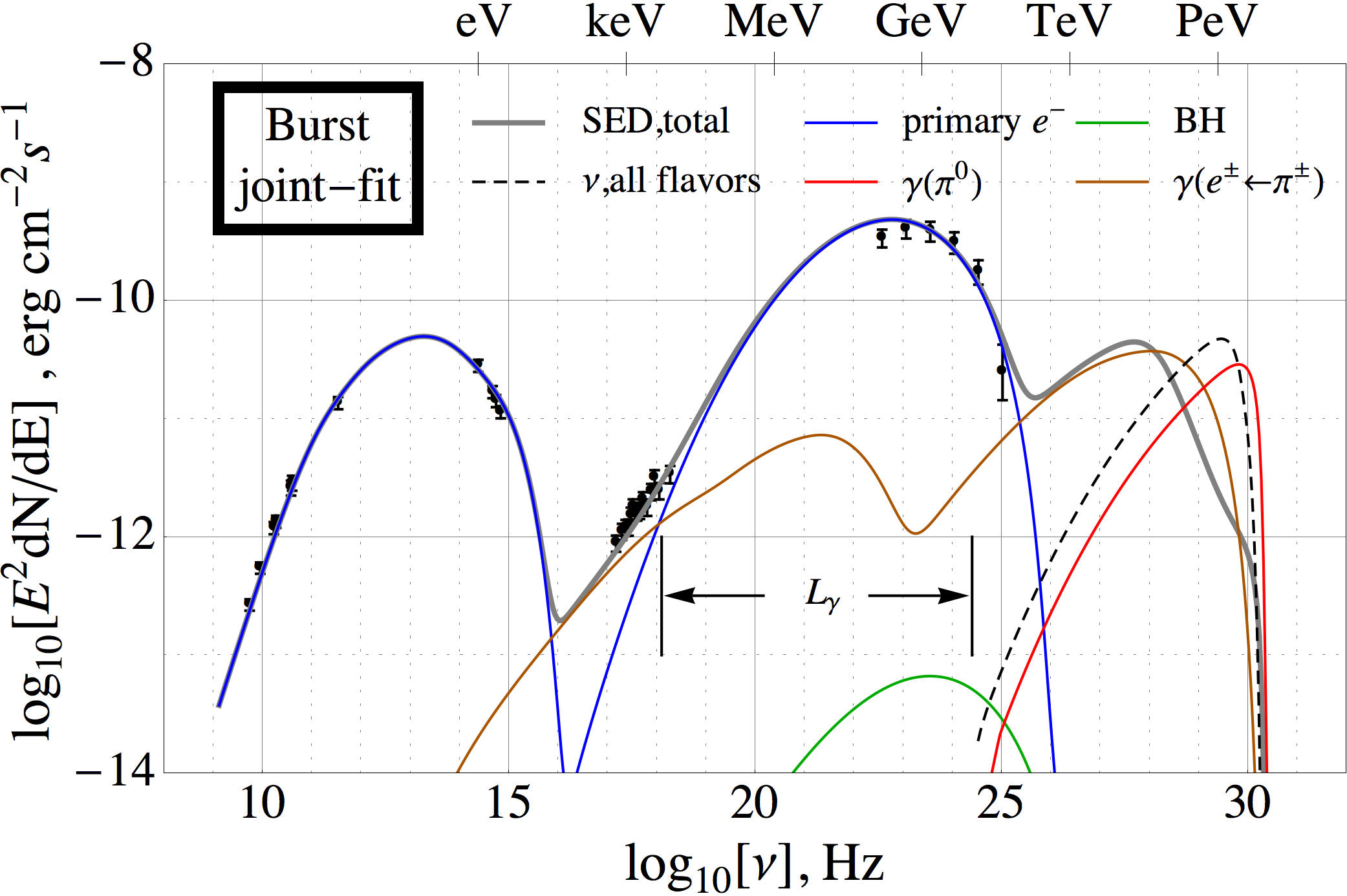} 
	\caption{ Components for the joint best-fit SED during burst phase. Gray: total SED; blue: emission from primary $e^{-}$; green: emission from pairs generated via Bethe-Heitler process; brown: $\gamma$-ray {\it injection} spectrum from $e^{\pm}$ pairs via $\pi^{\pm}$ decay. red: $\gamma$-ray {\it injection} spectrum from $\pi^{0}$ decay; black-dashed: neutrino, all flavors. The fractional contributions of these components to the $L_{\gamma}$ band, are $L_{ssc}=0.981,~L_{\pi}=0.0185,~L_{BH}=1.3\times10^{-4}$, respectively, where $L_{\gamma}$ is defined as the integrated luminosity between $10^{18.1}\sim10^{24.4} \mathrm{~Hz}$ (5 keV to 10 GeV). The bolometric neutrino to gamma-ray luminosity ratio is $L_{\nu}/L_{\gamma}=0.051$. The $\gamma\gamma$-absorption effect in the source becomes significant above $\sim100$ TeV, which is manifested in the suppression of the total SED in that energy band.  The data points here and in \Fig~\ref{fig:burst},\ref{fig:2lac} and \ref{fig:ic2yr} are provided by authors of K16 (processed from data by Fermi-LAT, Swift-XRT/UVOT, SMARTS, and the LBA, etc; see the supplementary material for the data analysis methods in K16).} 
	\label{fig:rates}
\end{figure*}


We find that both SED and neutrino fits independently prefer large baryonic loadings for the Burst phase, which may point towards a baryonically loaded burst. At the neutrino and joint best-fit points in \Fig~\ref{fig:burst} ($\times$,\ding{83}), the associated proton maximum energies are around $10$ PeV, which is consistent with the observed PeV neutrino event.  However, the maximal proton flux at the neutrino best-fit $\times$ is in tension with the X-ray data, which we will demonstrate below.

We show the SED for the SED best-fit (+), the joint best-fit (\ding{83}) and the neutrino best-fit ($\times$) in the left panel of \Fig~\ref{fig:burst}. We clearly observe that the SED is in tension with data in the X-ray energy range.  On the other hand, the SED is described reasonably well for the other two fit points. Note that the radio data are fitted from the extended emission region.

As a next step, we address the question whether the observed neutrino event can come from the Burst phase from PKS B1424-418, as reported in K16. One test of this hypothesis is energetics, \ie, in K16, the neutrino luminosity was directly related to the section of the SED we defined ``$L_{\gamma}$'' in the caption of \Fig~\ref{fig:rates}. One can easily see that this energy range is dominated by leptonic processes (1) in our model, contrary to what has been assumed in \Ref~K16 (for which the hadronic processes 2--4 would need to dominate that energy range). 

We list the fractional contributions to the SED of different physical interactions for the joint best-fit case of Burst phase in the caption of \Fig\ref{fig:rates} , where one can clearly read off that the SSC contribution dominates. We also show the neutrino-to-$\gamma$-ray ratio $L_\mathrm{\nu}/L_\mathrm{\gamma}$, which is order 5\% for the models fitting the SED. These numbers are to be interpreted as an additional ``theory'' correction factor in addition to those included in K16 (independent of the spectral correction in K16, which would reduce this number). They reflect the fact that hadronic processes only dominate in a small portion of the energy range marked ``$L_\mathrm{\gamma}$'' in \Fig\ref{fig:rates}, considering that the second hump cannot be dominated by hadronic processes.  Note that hadronic components contribute significantly to the SED outside the pre-defined energy range of $L_{\gamma}$ ($10^{18.1}\sim10^{24.4} \mathrm{~Hz}$, 5 keV to 10 GeV).  In K16, the predicted number of neutrino events in $1.0-2.0 \mathrm{~PeV}$ bin is 1.6, assuming that the entire second hump is generated from hadronic processes. From our model, this needs to be corrected by this factor of 0.05, arriving at $\sim 0.08$ events. Consistently, the neutrino spectrum from our numerical simulation (the joint best-fit case) predicts $0.094$ events in IceCube within the same energy bin.

\subsection{SED and Neutrino from 2LAC and IC-2yr Phase, and variation in Activity States}

\begin{figure*}[h]
	\centering
	\includegraphics[width=1.0\columnwidth]{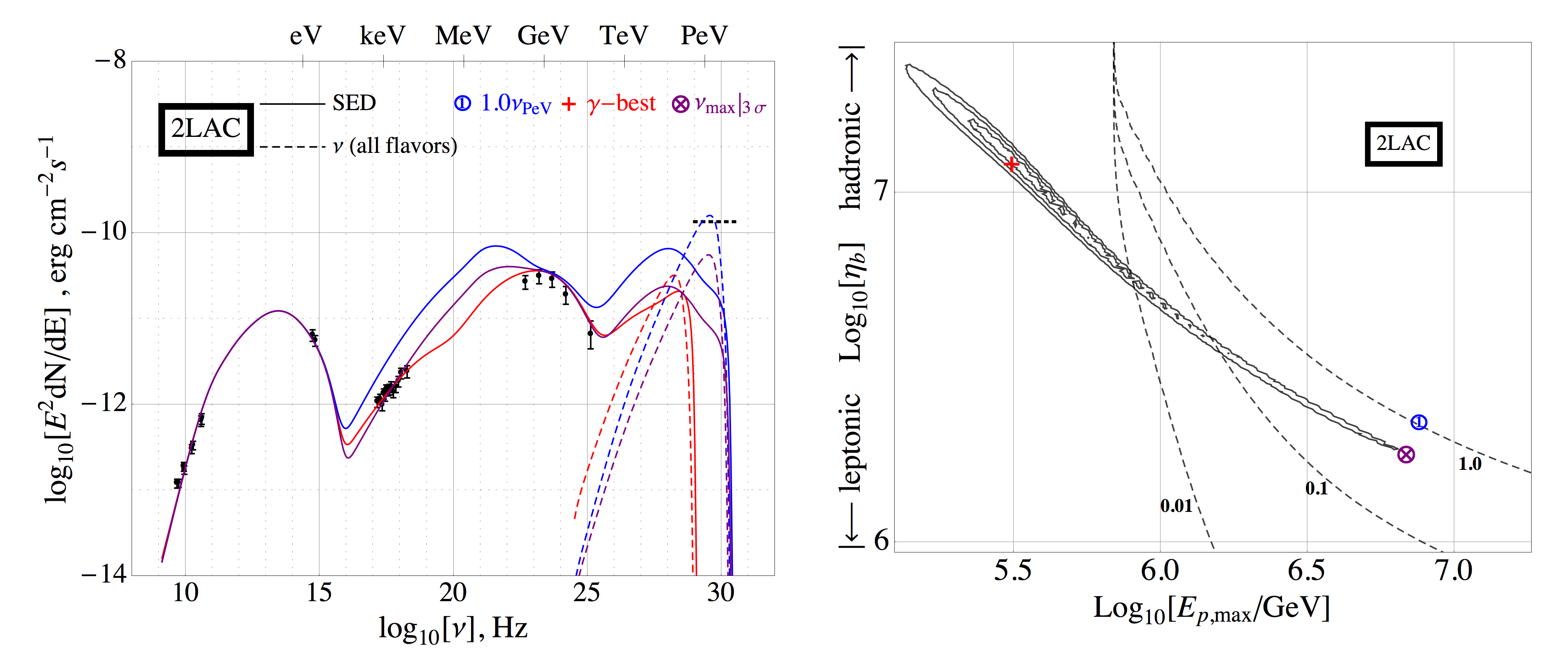} 
	\caption{Right panel: fitting to the SED, with $1-3\sigma$ confidence regions (boundaries in solid curves), and iso-neutrino event (between 0.5 and 2.4 PeV) contours (dashed). \ding{172} is the mark for maximum neutrino production within the $3\sigma$ region and \ding{83} is a representative point on the 1-neutrino event contour. Left panel: the corresponding SED (solid) and neutrino (dashed) spectra for those three marks on the right-panel. The short horizontal dashed-line is the neutrino flux level to expect one PeV-event in IC40+59 configuration during this phase.}
	\label{fig:2lac}
\end{figure*}

\begin{figure*}[h]
	\centering
	\includegraphics[width=1.0\columnwidth]{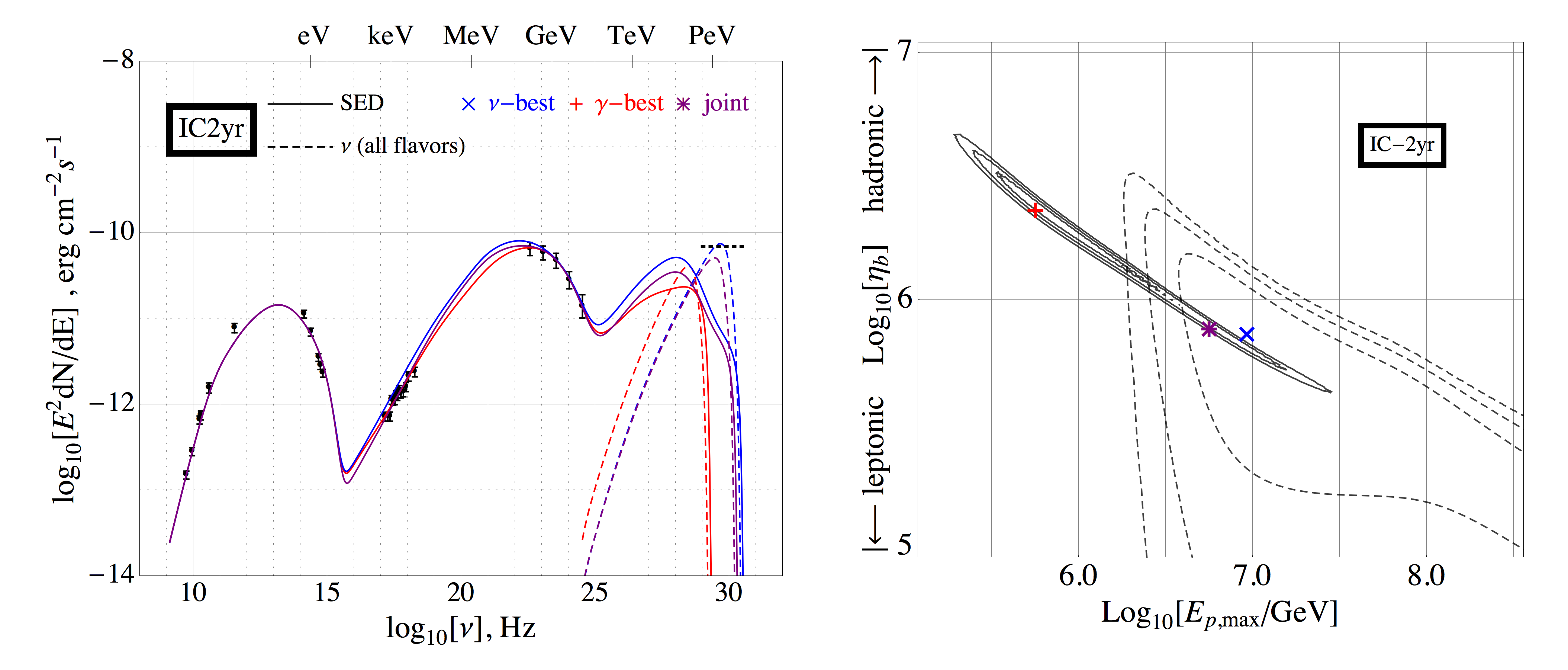}
	\caption{The fitting quality (right panel) and the corresponding SED + neutrino spectra (left panel) for the best-fit marks. The conventions for the curves and symbols are the same as those in \Fig~\ref{fig:burst}.}
	\label{fig:ic2yr} 
\end{figure*}


Let us now address if we can draw a self-consistent picture of the AGN blazar over time.  For that purpose, we independently fit the parameters for the three phases 2LAC, IC-2yr, and Burst in \Fig~\ref{illus}; these fits are shown in \Fig~\ref{fig:burst},\ref{fig:ic2yr} and \ref{fig:2lac}. For the IC-2yr phase, we predict the neutrino events under the same format as the one in the Burst phase, even though no neutrinos are detected during this phase. The purpose is to demonstrate that the correlation between the PeV neutrino event with the Burst phase, is weak. Indeed, the IC-2yr phase predicts up to a probability $P_{0,1,0}=5.7\%$ to produce the same observations in IceCube, compared to $P_{0,1,0}=3.2\%$ in the Burst phase (see \Tab~\ref{resulttable}). The relative probability is therefore $P_{0,1,0}(\mathrm{IC-2yr}):P_{0,1,0}(\mathrm{Burst})=1.6:1$. Here $P_{0,1,0}$ is defined as the joint probability to observe 0,1,0 neutrino events in the $0.5-1.6$, $1.6-2.4$ and $>2.4$ PeV energy bins, respectively, where in each bin, the expected number of neutrino events follows a Poisson distribution.

For the 2LAC phase, IceCube was operating with mainly IC40+59 configuration, compared to the full IC86 for the other phases. Here we show the contours of the predicted neutrino events of $>0.5$ PeV during this phase in the parameter space, from IC40+59 configuration. The best-fit point for the SED in \Fig~\ref{fig:2lac} is marked with ``$+$''; the maximum neutrino flux within the 3$\sigma$ region for the SED fit is marked by ``$\otimes$'', and a representative point on the $N_{\nu}=1$ contour is picked and marked with ``\ding{172}''.  Since the SED of the 2LAC phase has a lower photon flux than that of the IC-2yr or Burst phase, a lower target photon density in the source is implied.  In order to have a sufficiently large $p\gamma$ interaction rate to produce the right amount of X-rays and one neutrino event, the required proton-to-lepton luminosity ratio $\eta_{b}$ has to be very large: $\eta_{b} \gtrsim 10^{6}$. From that perspective it is not surprising that no neutrinos were observed during that phase.

\begin{table*}[tbp] 
	\centering
	\begin{tabular}{|c|ccc|ccc|ccc|} 
		\hline 
		Phase &	\multicolumn{3}{c|}{2LAC} & \multicolumn{3}{c|}{IC-2yr} & \multicolumn{3}{c|}{Burst} \\
		Time & \multicolumn{3}{c|}{2008.9--2010.9} & \multicolumn{3}{c|}{2010.5--2012.5} & \multicolumn{3}{c|}{2012.6 -- 2013.3} \\
		\hline
		$R_\mathrm{blob}/{\rm cm}$	&  \multicolumn{9}{c|}{$7.5\times 10^{17}$} \\
		$\Gamma_\mathrm{bulk}$& 	\multicolumn{9}{c|}{35} \\
		\hline
		$B^{\prime}/{\rm mG}$ & \multicolumn{3}{c|}{2.5} & \multicolumn{3}{c|}{2.0} & \multicolumn{3}{c|}{2.5} \\
		$L_\mathrm{e,inj}/L_\mathrm{edd}$ & \multicolumn{3}{c|}{$6.7\times 10^{-5}$} & \multicolumn{3}{c|}{$1.2\times 10^{-4}$} &\multicolumn{3}{c|}{$3.2\times 10^{-4}$}  \\
		$\gamma_\mathrm{e,min}^{\prime}$ &\multicolumn{3}{c|}{$1.8\times 10^{3}$} & \multicolumn{3}{c|}{$1.8\times 10^{3}$} &\multicolumn{3}{c|}{$2.2\times 10^{3}$}  \\
		$\gamma_\mathrm{e,max}^{\prime}$ &\multicolumn{3}{c|}{$1.3\times 10^{5}$} & \multicolumn{3}{c|}{$8.9\times 10^{4}$} &\multicolumn{3}{c|}{$1.0\times 10^{5}$}  \\
		$\alpha_\mathrm{e,idx}^{\prime}$ & \multicolumn{3}{c|}{-2.2} & \multicolumn{3}{c|}{-2.2} &\multicolumn{3}{c|}{-1.8} \\ 
		\hline
		$\alpha_\mathrm{p,idx}^{\prime}$ &\multicolumn{9}{c|}{-2.0} \\ \hline \hline
		Fit symbol  &		$+$ & $\otimes$ & \ding{172} & $+$	&	\ding{83} &	$\times$	& $+$	&	\ding{83} &	$\times$	 \\
		Fit  & SED & \Tab~\ref{paratable} & \Tab~\ref{paratable}  & SED & joint & $\nu$ & SED & joint & $\nu$\\
		\hline
		$\eta_\mathrm{b}/10^{6}$                 & 11.5 & 1.8 & 2.2 & 2.3  & 0.76 & 0.72 & 0.33	& 0.10 & 0.20 \\
		$E_\mathrm{p,max}^\mathrm{ob}/{\rm PeV}$ & 0.37 & 7.5 & 8.3 & 0.68 & 6.8  & 11.2 & 0.68 & 7.5  & 12.3 \\
		\hline 
		$N_{\nu}, 0.5-1.6~ \mathrm{PeV}$  					& 0    & 0.35  & 0.74 & 0.0090 & 0.62  & 0.79 & 0.0045 & 0.21  & 0.76 \\
		$N_{\nu}, 1.6-2.4~ \mathrm{PeV}$  					& 0    & 0.082 & 0.18 & 0      & 0.13  & 0.22 & 0      & 0.042 & 0.22 \\ 
		$N_{\nu}, ~~~~>2.4~ \mathrm{PeV}$ 					& 0    & 0.037 & 0.10 & 0      & 0.044 & 0.21 & 0      & 0.018 & 0.24 \\
		$P_{0,1,0}$, \% 	& -    & -     & -    & 0      & 5.7   & 6.5  & 0      & 3.2   & 6.4  \\ 
		\hline 
		photon SED $\chi^{2}/\text{d.o.f.}$ 				& 0.83 & 1.36  & 86.9 & 1.82   & 1.94  & 7.10 & 1.38   & 1.40  & 171  \\
		\hline 
	\end{tabular}
	\caption{Parameters and the expected number of neutrino events in IceCube for the best-fit models during each phase. The symbol $P_{0,1,0}$ next to the bottom row is the joint-probability to observe 0,1,0 neutrino events in the three energy bins above. See \Tab~\ref{paratable} for the definition of symbols and notations.} 
	\label{resulttable} 
\end{table*}

The X-ray band, seen as the gap between the two humps in the SED, is a mixture of leptonic and hadronic components, as shown in Fig.\ref{fig:rates}. Hadronic secondary-emission depends on the detailed processing of the electromagnetic cascades in the source, mainly via the following channels: $p\gamma \rightarrow \pi^{0} \rightarrow \gamma\gamma \rightarrow e^{\pm}$, $p\gamma \rightarrow \pi^{\pm} \rightarrow \mu^{\pm}\rightarrow {e^{\pm}}$, and $p\gamma \rightarrow p+e^{\pm}$.  In the $\pi^{0}$ channel, the emission predominantly comes from the first generation of pairs. The source itself becomes optically thick for $\gamma$-rays above roughly 100 TeV.  It can be further estimated by \cite{SteckerEBL12} that the spectrum above $\sim100$ GeV will suffer from additional suppression due to absorption by EBL, but this effect does not play a significantly role in our fitting.

We list the detailed best-fit parameters and expected number of neutrino events during each phase in \Tab~\ref{resulttable}. We notice that the parameters for the magnetic fields, minimum and maximum energies for $e^{-}$ and $p$ injections are very similar. The SEDs evolving from the earliest 2LAC phase to IC-2yr and finally the Burst phase can be simply reproduced by an increase of the $e^{-}$ versus ${p}$ injection rate with time, and a slightly hardening effect on the $e^{-}$ injection spectrum. 

The proton-to-lepton luminosity ratio required to fit the neutrino result is large, of the order of $10^{5}$--$10^{6}$. This value is coincidently similar to that found for other blazars, such as \cite{Diltz15,Petropoulou1501}.  In our case, on one hand the p$\gamma$ interaction rate is lower than those in the above cases and more protons are needed to achieve the same level of $\gamma-$rays; on the other hand, compared to their LH$-\pi$ models, we only need a subdominant hadronic contribution to $\gamma-$rays, which lowers the requirement for proton luminosity.  The physical proton-injection luminosities needed for these values are around a few times the Eddington luminosity.  This requirement can be alleviated if we assume a lower escape rate for protons, \eg\ a magnetic configuration that can trap the protons longer in the blob. In our model, due to the low $p\gamma$ efficiency, most protons simply escape without having a $p\gamma$ interaction and the interactions cause no visible effects on the proton spectrum. Therefore, the required $L_\mathrm{p,inj}$ scales linearly with the escape rate.

\section{Summary and conclusions}
\label{conclusion}

A potential correlation between the PeV neutrino event ``Big Bird'' by IceCube in 2012 and the Burst phase of FSRQ PKS B1424-418 was reported by \cite{Kadler16} (K16). That analysis relies on the frequently used assumption $L_\mathrm{\gamma}\approx L_\mathrm{\nu}$. In this paper we have revisited this important relationship with a self-consistent one-zone emission model, with SEDs measured from three phases of the source: 2LAC (2008.8-2010.9), IceCube 2-year (IC-2yr, 2010.5-2012.5) and Burst phase (2012.6-2013.3).  Those different SEDs can be interpreted in our results as variations on blazar properties over time.  We have also studied the parameter values needed to attribute to the neutrino event, which was observed in the IC-2yr phase and within that, the Burst phase. 

We found that the ``conventional'' hadronic model (high energy hump via hadronic processes; the ``LH-$\pi$'' or ``LH-psyn'' model) does not work for this blazar, which is different from many other blazars, such as the well-studied Mrk 421 \cite{Greek14, Petropoulou1603}, for which the high-energy hump of SED can be fully accounted for by hadronic processes. Therefore, the leading contribution to the SED of PKS B1424-418 must be leptonic.  We have, however, shown that a subdominant hadronic contribution can be present. In terms of neutrino production, it is therefore an interesting question how much baryonic loading can be tolerated, even in an SSC model, without affecting the shape of the SED.

We have found that the observed one neutrino event reported by K16 during the Burst phase is in tension with the SED fits, whereas up to $\sim0.3$ above-PeV events are acceptable. The probability to produce 0, 1, and 0 neutrinos in the $0.5-1.6$, $1.6-2.4$, and $>2.4$ PeV bins, respectively, as observed by IceCube, is up to $5.7\%$ in IC-2yr and $3.2\%$ in Burst phase. This suggests a chance coincidence of the observed PeV-event with the Burst phase of PKS B1424-418, since there is a even higher probability for it to occur during the IceCube 2-year observation than the Burst phase. Our predicted event rates are consistent with K16 if their expectation is corrected for small fraction of the photon energy flux in the second peak comes from hadronic processes.  This means that an additional ``theory correction factor'' has to be applied if the gamma-ray and neutrino fluxes are to be correlated, which can only come from self-consistent models. 

We have also demonstrated from independent simulations of the different phases of the blazar that a self-consistent time-dependent evolution picture can be drawn, meaning that the different phases can be described by similar parameters. We noted that the active state (Burst phase) can be achieved simply by increasing the injection rate of electrons and protons. It is the data in the X-ray range which sets constraints on the baryonic loading.

While our results apply to PKS B1424-418 specifically, we emphasize the importance of a self-consistent description of the SED as opposed to to generic approaches relating the gamma-ray and neutrino luminosities. The studied blazar serves as a counter-example for the direct correlation of these, which means that this assumption does not hold in general and has to be used with care. We have also presented an example in a quantitative way on how the neutrino observation, even with one candidate event, is able to break the degeneracies of blazar models.

\acknowledgements
We thank the authors of K16, especially M.Kadler and F.Krauss for providing the data of PKS B1424-418.
SG and MP acknowledge support by the Helmholtz Alliance for Astroparticle Physics HAP funded by the Initiative and Networking Fund of the Helmholtz Association.  WW acknowledges funding from the European Research Council (ERC) under the European Union's Horizon 2020 research and innovation programme (Grant No. 646623).

\appendix


\section{Analytical constraints}
\label{appendix:ana}

For brevity, all quantities are expressed in the blob-comoving frame, unless annotated by a superscript ``$ob$''. 

The characteristic energy of the proton is constrained by the observed PeV neutrino event, as
\begin{equation}
	\gamma_\mathrm{p,char}\sim E_\mathrm{\nu}m_\mathrm{p}c^{2}/K_\mathrm{\nu}\,
	\label{equ:characteristicprotonE}
\end{equation}
where $E_\mathrm{\nu}=E_\mathrm{\nu}^\mathrm{ob}(1+z)/\Gamma$ is the energy of the neutrino, and $K_\mathrm{\nu}\sim0.05$ is the characteristic ratio of $E_{p}:E_{\nu}$ from the $p\gamma$ interaction. The $p\gamma$ event rate, per physical volumn, estimated via the $\Delta^{+}(1232)$ resonance, is
\begin{equation}
	\dot{N}_\mathrm{p\gamma}\sim c\sigma_\mathrm{p\gamma}f_\mathrm{ph}(\varepsilon_\mathrm{p\gamma,t})f_\mathrm{p}(\gamma_\mathrm{p,char})
	\label{equ:pgammareactionrate}
\end{equation}
where $\sigma_\mathrm{p\gamma}=5.0\times10^{-28} {\rm~cm^{2}}$ is the cross-section, $f_{i}(E_{i})\equiv E_{i}\dfrac{dN_{i}}{dE_{i}}$ is the number density of particle $i$ around $E_{i}$ ($f_\mathrm{ph}$ is the photon density derived from the observed SED), $\varepsilon_\mathrm{p\gamma,t}\sim \varepsilon_\mathrm{p\gamma,th}\gamma_\mathrm{p,char}^{-1}$ is the target photon energy and $\epsilon_\mathrm{p\gamma,th}\sim0.3{\rm~GeV}$ is the positon of the $\Delta$ resonance in the energy axis.

The injection rate in terms of the energy density of all pions due to $p\gamma$ interaction is 
\begin{equation}
	\dot{u}_{\pi}=\dot{N}_\mathrm{p\gamma}E_{p}K_\mathrm{p\gamma}\,
	\label{equ:udotpi}
\end{equation}
where $K_\mathrm{p\gamma}\sim0.2$ is the average inelasticity for the proton in $p\gamma$ interaction and $E_\mathrm{p,char}$ and $\dot{N}_\mathrm{p\gamma}$ are computed from \equ{characteristicprotonE} and \equ{pgammareactionrate}, respectively.

The luminosity of the second hump is given by the synchrotron photons from the secondaries as a result of pion decay
\begin{equation}
	L_\mathrm{ph,2}\sim \alpha_\mathrm{fs}\varepsilon_{2}f_\mathrm{ph}(\epsilon_{2})V_\mathrm{blob}\sim K_{\pi\rightarrow e}\dot{u}_{\pi}V_\mathrm{blob}
	\label{equ:L2normalization}
\end{equation}

where $K_{\pi\rightarrow e}\sim 1/8$ (or higher, up to 5/8 if all $\gamma-$rays from $\pi^{0}$ decays are absorbed {\it in situ}.) is the fraction of energy transferred to $e^{\pm}$ from $\pi$ decays and $\alpha_\mathrm{fs}=4c/(3R)$ is the escape rate of the photons in the free-streaming case. Combining \equ{udotpi}, \equ{L2normalization} we obtain the constraint on the steady-state proton number density around the energy $E_\mathrm{p,char}$
\begin{equation}
	f_\mathrm{p}(\gamma_\mathrm{p,char})=\dfrac{\alpha_\mathrm{fs}}{c\sigma_\mathrm{p\gamma}}\dfrac{\varepsilon_{2}}{E_\mathrm{p,char}}\dfrac{f_\mathrm{ph}(\varepsilon_{2})}{f_\mathrm{ph}(\varepsilon_\mathrm{p\gamma,t})}\dfrac{1}{K_\mathrm{p\gamma}K_{\pi\rightarrow e}} \,
	\label{equ:fpchar}
\end{equation}
together with the synchrotron peak energy from the secondaries in the unit of $m_{e}c^{2}$
\begin{equation}
	\varepsilon_\mathrm{p\gamma,pk}=F_\mathrm{b}\gamma_\mathrm{p\gamma,e}^{2}
	\label{equ:pgammapeak}
\end{equation}
where $\gamma_\mathrm{p\gamma,e}=r_\mathrm{m}^{-1}K_\mathrm{p\gamma}K_{\pi\rightarrow e}\gamma_\mathrm{p,char}$.

The injection rate of the energy density of pairs via Bethe-Heitler effect is estimated as
\begin{equation}
	\dot{u}_\mathrm{e,bh}=c\sigma_\mathrm{bh}K_\mathrm{bh}f_\mathrm{ph}(\varepsilon_\mathrm{bh,t})f_\mathrm{p}(\gamma_{p})E_\mathrm{p,char}\,
	\label{equ:uebhdot}
\end{equation}
where $\sigma_\mathrm{bh}$ is the cross-section of Bethe-Heitler process and $K_\mathrm{bh}$ is the inelasticity. Both are dependent on the energy of the incident photon in the proton rest frame; however, the product of them has a peak value of 
\begin{equation}
	\sigma_\mathrm{bh}K_\mathrm{bh}\sim1.5{\rm~\mu b}
	\label{equ:effectivebhcrosssection}
\end{equation}
at the proton-rest-frame photon energy 
\begin{equation}
	\varepsilon_\mathrm{bh,r}\sim\gamma_\mathrm{p,char}\varepsilon_\mathrm{bh,t}\sim25 m_{e}c^{2}
	\label{equ:bhcharenergy}
\end{equation}.

Substituting \equ{fpchar} into the equation above, we obtain the luminosity of the synchrotron photons from those pairs, as
\begin{equation}
	\varepsilon_\mathrm{bh,pk}f_\mathrm{bh}(\varepsilon_\mathrm{bh,pk})=\dfrac{\sigma_\mathrm{bh}}{\sigma_\mathrm{p\gamma}}\dfrac{K_\mathrm{bh}}{K_\mathrm{p\gamma}K_{\pi\rightarrow e}}\dfrac{f_\mathrm{ph}(\varepsilon_\mathrm{bh,t})}{f_\mathrm{ph}(\varepsilon_\mathrm{p\gamma,t})}\varepsilon_{2}f_\mathrm{ph}(\varepsilon_{2})\,
	\label{equ:nufnubh}
\end{equation} 
where the peak energy of the synchrotron photons from those pairs are
\begin{equation}
	\varepsilon_\mathrm{bh,pk}=F_\mathrm{b}\gamma_\mathrm{e,bh}^{2}\,
\end{equation}
in the unit of $m_{e}c^{2}$ with $F_\mathrm{b}=B/B_\mathrm{crit}$ defined under \equ{Rsyn} and characteristic energy of those pairs being $\gamma_\mathrm{e,bh}=r_{m}^{-1}K_\mathrm{bh}\gamma_\mathrm{p,char}/2$, where $r_{m}\equiv m_{e}/m_{p}$ is the $e^{-}$ to $p$ mass ratio and $K_\mathrm{bh}=2\chi_\mathrm{BH}(\varepsilon_\mathrm{bh,r})\sim0.25r_\mathrm{m}$ is numerically calculated from \equ{chiBH}.

The proton synchrotron has a peak energy of 
\begin{equation}
	\varepsilon_\mathrm{psyn,pk}=r_\mathrm{m}F_\mathrm{b}\gamma_\mathrm{p,char}^{2}
	\label{equ:varepsilonpsyn}
\end{equation}
in the unit of $m_{e}c^{2}$. The injection rate of energy density of proton-synchrotron photons is 
\begin{equation}
	\dot{u}_\mathrm{psyn}=\dfrac{4}{3}r_\mathrm{m}^{3}c\sigma_{T}u_\mathrm{b}\gamma_\mathrm{p,char}^{2}f_\mathrm{p}(\gamma_\mathrm{p,char})m_\mathrm{p}c^{2}\ .
\end{equation}

By substituting $f_\mathrm{p}$ with the expression from \equ{fpchar} we get 
\begin{equation}
	f_\mathrm{psyn}(\varepsilon_\mathrm{psyn,pk})=\dfrac{4}{3}r_\mathrm{m}u_\mathrm{b}F_\mathrm{b}^{-1}K_\mathrm{p\gamma}^{-1}K_{\pi\rightarrow e}^{-1}\dfrac{\sigma_{T}}{\sigma_{p\gamma}}\dfrac{\varepsilon_{2}}{E_\mathrm{p,char}}\dfrac{f_\mathrm{ph}(\varepsilon_{2})}{f_\mathrm{ph}(\varepsilon_{p\gamma,t})}\ .
	\label{equ:fpsyn}
\end{equation}

Since the scattering is mainly in Thomson regime (\equ{KN}) the $Y_\mathrm{ssc}$ parameter is estimated by the energy density ratio of low energy photons to magnetic field
\begin{equation}
	Y_\mathrm{ssc}\approx u_\mathrm{phot}/u_\mathrm{b}\approx \varepsilon_{1}f_\mathrm{ph}(\varepsilon_{1})/u_\mathrm{b}\ .
	\label{equ:ysscestimate}
\end{equation}

The constraints for \Fig~\ref{lhpimodel} are then calculated from the following equations: \ref{constraint:1}:\equ{fpsyn}; \ref{constraint:2}:\equ{ysscestimate}; \ref{constraint:3}:\equ{pgammapeak}; \ref{constraint:4}:\equ{nufnubh}

\section{Kinematic equations}
\label{appendix:kin}

We simulate time-dependent particle spectra for the above particle species. The kinematics of those particles are described by the following set of coupled integro-differential equations:

\begin{equation} 
	\partial_{t}n(\gamma,t)=-\partial_{\gamma}\{{\dot{\gamma}({\gamma,t})n(\gamma,t)-\partial_{\gamma}[D(\gamma,t) n(\gamma,t)]/2}\}-\alpha(\gamma,t) n(\gamma,t)+Q(\gamma,t)
	\label{equ:overalllinearform}
\end{equation}

where the $n(\gamma,t)$ is the differential number density of the particle and the total number of particles equals $N=\int dV\int d\gamma n(\gamma)$. $\gamma$ denotes the Lorentz factor of $e^{+}$, $e^{-}$ or the dimensionless energy of the photon $\gamma_{f}=E_{f}/m_{e}c^2$.

On the right hand side of the equation, $Q(\gamma,t)$ is the source term, representing the generation and the injection rate of the particle. For the process of $a+b\rightarrow c+d$, the generation rate of particle specie $c$ can be written as an integration over the parent particle population $b$:

\begin{equation}
	Q_{c}(\gamma_{c})=\int R(c\leftarrow b)n_{b}(\gamma_{b})d\gamma_{b}
	\label{eqn:Qc}
\end{equation}

where the integration kernal $R(c\leftarrow b)$ depends on a further layer of integration

\begin{equation}
	R(c\leftarrow b)=\int R(c\leftarrow a,b)n_{a}(\gamma_{a})d\gamma_{a} 
\end{equation}

and $R(c\leftarrow a,b)$ is the differential cross-section of generating the particle $c$ with $\gamma_{c}$, averaged over the reaction angle between the incident particle $a$ and $b$ in the lab frame.

\begin{equation}
	R(c\leftarrow a,b)=\frac{c}{2}\int (1-\mu)\frac{d\sigma}{d\gamma_{c}d\mu}(\gamma_{c},\gamma_{b},\gamma_{a},\mu)d\mu
\end{equation}

where $\mu\equiv cos{\theta}$ and $\theta$ is the reaction angle between $b$ and $c$.

In a process like $b+a\rightarrow b+c$, with the particle $b$ reappearing with a different energy, the particle $b$ on the left-hand side is nevertheless treated as ``annihilated'' here. The dissapearance rate for $b$, is therefore

\begin{equation}
	\alpha(\gamma_{b})=\int R(a,b)n_{a}(\gamma_{a})d\gamma_{a}
\end{equation}

where $R(a,b)$ is

\begin{equation}
	R(a,b)=\int R(c\leftarrow a,b)d\gamma_{c}
\end{equation}. 
 
 and the re-appeared $b$ after the reaction is treated as a new particle. The re-appearance rate can be obtained by eqn.\ref{eqn:Qc}.


 \begin{table*}[h]
	\centering
	\begin{tabular}{|c|c|c|c|c|c|c|c|}
		\hline 
				& injection & escape & synchrotron & inverse Compton & $\rm \gamma\gamma\leftrightarrow e^{\pm}$ & Bethe-Heitler & $p\gamma$ \\
 		\hline
 		$\rm e^{-}$ & $\rm Q_{e,inj}$ & $\rm \alpha_{e,esc}$ & $\rm \dot{\gamma}_{e,syn},~D_{e,syn}$ & $\rm \dot{\gamma}_{e,IC},~D_{e,IC},~\alpha_{e,IC},~Q_{e,IC}$ & $\rm \alpha_{e,pa},~Q_{e,pp}$ & $\rm Q_{BH}$ & $\rm Q_{e,p\gamma}$ \\
		\hline
		$\rm e^{+}$ & -- & $\rm \alpha_{e,esc}$ & $\rm \dot{\gamma}_{e,syn},~D_{e,syn}$ & $\rm \dot{\gamma}_{e,IC},~D_{e,IC},~\alpha_{e,IC},~Q_{e,IC}$ & $\rm \alpha_{e,pa},~Q_{e,pp}$ & $\rm Q_{BH}$ & $\rm Q_{e,p\gamma}$ \\
		\hline
		$\rm \gamma$ & -- & $\rm \alpha_{f,esc}$ & $\rm \alpha_{f,ssa},~Q_{f,syn}$ & $\rm \alpha_{f,IC},~D_{f,IC}$ & $\rm \alpha_{f,pp},~Q_{f,pa}$ & $\rm \alpha_{f,BH}$ & $\rm \alpha_{f,p\gamma},~Q_{f,p\gamma}$ \\
		\hline
		$\rm p$ & $\rm Q_{p,inj}$ & $\rm \alpha_{e,esc}$ & $\rm \dot{\gamma}_{p,syn},~D_{p,syn}$ & $\rm \dot{\gamma}_{p,IC}~D_{p,IC},~\alpha_{p,IC},~Q_{p,IC}$ & -- & $\rm \dot{\gamma}_{p,BH},~D_{p,BH}$ & $\rm \alpha_{p,p\gamma},~Q_{p,p\gamma}$ \\
		\hline
		$\rm n$ & -- & $\rm \alpha_{f,es}$ & -- & -- & -- & -- & $\rm \alpha_{n,p\gamma},~Q_{n,p\gamma}$ \\
		\hline
		$\rm \nu$ & -- & $\rm \alpha_{f,es}$ & -- & -- & -- & -- & $\rm Q_{\nu,p\gamma}$ \\
		\hline
  \end{tabular}
	\caption{List of coefficients.}
	\label{AppendixTable}
\end{table*}


\subsection{inverse Compton}


If we consider the relativistic electrons only, e.g. $\gamma_{e}>10$, the differential cross-section (expressed in the unit of $\sigma_{T}$) can be much simplified under the head-on collision approximation \citep{DermerBook09} (DM09)

\begin{equation}
	\frac{d^{2}\sigma_\mathrm{IC}}{d\gamma_{f,o}d\mu}\left(\gamma_{e,o},\gamma_{e,i},\gamma_{f,i},\mu\right)=
	\frac{3}{8}\frac{1}{\gamma_{e,i}x^{3}} \left[ yx^{2}+1+2x+\frac{1}{y}\left(x^2-2x-2\right)+\frac{1}{y^2} \right]
	H\left( y-\frac{1}{1+2x} \right) H\left(1-\frac{x}{2\gamma_{e,i}^2}-y\right)
\end{equation}

where $x=\gamma_{e,i}\gamma_{f,i}(1-\mu)$ is the photon energy in the electron rest frame and $y=1-\gamma_{f,o}/\gamma_{e,i}$. The subindex $i$($o$) stands for the incoming(outgoing) particle and $e$($f$) represents electron(photon). $\gamma_{e}$ and $\gamma_{f}$ are dimensionless energy units defined by $\gamma_{e}=E_{e}/m_{e}c^2$ and $\gamma_{f}=E_{f}/m_{e}c^2$.

\begin{equation}
	\begin{aligned}
		& R_\mathrm{IC}\left(\gamma_{f,o}\leftarrow\gamma_{e,i},\gamma_{f,i}\right) 
			=
				\dfrac{c}{2}\int(1-\mu)\dfrac{d^{2}\sigma_{IC}}{d\gamma_{f,o}d\mu} \left(\gamma_{e,o},\gamma_{e,i},\gamma_{f,i},\mu \right) d\mu \\
		&	= 
				\begin{cases}c
					\dfrac{ 6(1-u)\dfrac{u}{v}\ln\left[\dfrac{u}{2v(1-u)}\right] - 3\left( \dfrac{u}{v}-2+2u \right) \left(\dfrac{u}{v}+1+\dfrac{u^{2}}{2}-u\right)}
					{4v\gamma_{e,i}(u-1)^{3}},	& {\rm~for~} \dfrac{u}{2\gamma_{e,i}^{2}}<v<\dfrac{2u}{1+2u} {\rm ~and~} u<2\gamma_{e,i}^{2}-1/2\\
					0							& {\rm otherwise}
				\end{cases}
	\end{aligned}
\end{equation}

where $u=2\gamma_{e,i}$, $v=\gamma_{f,o}/\gamma_{e,i}$.

In the Thomson scattering limit, it reduces to 
\begin{equation}
	R_\mathrm{IC}(\gamma_{f,o}\leftarrow\gamma_{e,i},\gamma_{f,i})=\dfrac{\ln(w/2)-(w/2-1/2-w^{-1})}{2\gamma_{f,o}/3}
\end{equation}

where $w=u/v=2\gamma_{e,i}^{2}\gamma_{f,i}/\gamma_{f,o}$.

The total reaction rate, averaged over the angle $\mu$, is

\begin{equation}
	\begin{aligned}
		R_\mathrm{IC}(\gamma_{f,i},\gamma_{e,i})	&	 = \int R_\mathrm{IC}(\gamma_{f,o}\leftarrow\gamma_{e,i},\gamma_{f,i})d\gamma_{f,o} \\
													&	 = \dfrac{3c}{8u^{3}(1+2u)}\left[
														-2u(4+9u+u^{2})+(4+u)(1+2u)^{2}\ln(1+2u)+4u(1+2u){\rm Li}(2,-2u)
													\right]
	\end{aligned}
\end{equation}

which has the asymptotic forms as

\begin{equation}
	R_\mathrm{IC}(\gamma_{f,i},\gamma_{e,i})=
		\begin{cases}
			c\left(1-\dfrac{4u}{3}\right), & {\rm~for~} u<<1 {\rm~(Thomson)} \\
			c\dfrac{3}{4u}\left[\ln\left(\dfrac{u}{2}\right)-\dfrac{1}{2}\right], & {\rm~for~} u>>1 {\rm~(Klein-Nishina)}
		\end{cases} \ .
\end{equation}

${\rm Li}(2,x)$ is the dilogarithm defined as
\begin{equation}
	{\rm Li}(2,z)=-\int_{0}^{z}\dfrac{\ln(1-t)}{t}dt
\end{equation}

In the Thomson regime, after scattering, the electron loses a tiny fraction of energy, and therefore the function $R_{IC}(\gamma_{e,o}\leftarrow\gamma_{f,i}\gamma_{e,i})$ is highly peaked around $\gamma_{e,i}$ which the numerical grid is unable to resolve. Here we use the differential terms to account for this effect in the numerical computation, up to the second order: (VP09)

\begin{equation}
	-\partial_{\gamma}\left\{\dot{\gamma}_{\!_\mathrm{IC}} n(\gamma_{e},t)-\partial_{\gamma}\left[D_{\!_\mathrm{IC}}(\gamma_{e},t)n(\gamma_{e},t)\right]  \right\}
	= -\alpha_{\!_\mathrm{IC}}(\gamma_{e},t)n(\gamma_{e},t)+Q_{\!_\mathrm{IC}}(\gamma_{e},t)
	\label{equ:continuousloss}
\end{equation}

Using the moment expansion (equations C11,C12,C18,C19 of VP09,) they are expressed as
\begin{equation}
	\dot{\gamma}_{\!_\mathrm{IC,e}}(\gamma_{e})=\int\gamma_{f}(\Psi_{1}-\Psi_{0})n_{f}(\gamma_{f})d\gamma_{f}\,
	\label{equ:gammaic}
\end{equation}
\begin{equation}
	D_{\!_\mathrm{IC,e}}(\gamma_{e})=\int\gamma_{f}^{2}(\Psi_{2}-2\Psi_{1}+\Psi_{0})n_{f}(\gamma_{f})d\gamma_{f}
	\label{equ:dic}
\end{equation}
where these moments are given by \cite{NP94},
\begin{equation}
	\begin{aligned}
		\Psi_{0}(\gamma_{f},\gamma_{e}) & \approx 1 - \dfrac{2}{3}\left(4\gamma_{e}^{2}-1\right)\gamma_{f}\gamma_{e}^{-1} + \dfrac{26}{5}\left(2\gamma_{e}^{2}-1\right)\gamma_{f}^{2}\\
		\Psi_{1}(\gamma_{f},\gamma_{e}) & \approx \dfrac{1}{3}\left(4\gamma_{e}^{2}-1\right) - \dfrac{1}{5}\left(42\gamma_{e}^{4}-29\gamma_{e}^{2}+2\right)\gamma_{f}\gamma_{e}^{-1} + \dfrac{1}{25}\left(1176\gamma_{e}^{4}-1147\gamma_{e}^{2}+206\right)\gamma_{f}^{2}\\
		\Psi_{2}(\gamma_{f},\gamma_{e}) & \approx \dfrac{1}{15}\left(42\gamma_{e}^{4}-34\gamma_{e}^{2}+7\right) - \dfrac{4}{75}\left(528\gamma_{e}^{6}-618\gamma_{e}^{4}+172\gamma_{e}^{2}-7\right)\gamma_{f}\gamma_{e}^{-1}\\
		& + \dfrac{1}{525}\left(109120\gamma_{e}^{6}-158856\gamma_{e}^{4}+63677\gamma_{e}^{2}-6066\right)\gamma_{f}^{2}\\
	\end{aligned}
\end{equation}


\subsection{Pair production and annihilation}


For the pair production process, $\gamma_{f,1}+\gamma_{f,2}\rightarrow \gamma_{e,1}+\gamma_{e,2}$, two photons with dimensionless energies $\gamma_{f,i}=E_{f,i}/m_{e}c^{2},~(i=1,2)$ are annihilated and an electron-positron pair is produced with Lorentz factors $\gamma_{e,1},\gamma_{e,2}$, respectively. Obviously, for energy conservation we have $\gamma_{f,1}+\gamma_{f,2}=\gamma_{e,1}+\gamma_{e,2}$. We use the treatment and expressions from VP09:

\begin{equation}
	Q_\mathrm{pp}(\gamma_{e,1})=c\int_{\gamma_{f,1}^{*}}^{\infty} n(\gamma_{f,1})d\gamma_{f,1} \int_{\gamma_{f,2}^{*}}^{\infty} R_{pp}(\gamma_{e,1}\leftarrow \gamma_{f,1},\gamma_{f,2})n(\gamma_{f,2})d\gamma_{f,2}
	\label{equ:Qpp}
\end{equation}
where 
\begin{equation}
	R_\mathrm{pp}(\gamma_{e,1}\leftarrow\gamma_{f,1},\gamma_{f,2})=-\gamma_{f,1}^{-2}\gamma_{f,2}^{-2}\left[S(\gamma_{e,1},\gamma_{f,2},\gamma_{f,1},w_{U})-S(\gamma_{e,1},\gamma_{f,2},\gamma_{f,1},w_{L})\right]/4
\end{equation}
in which,
\begin{equation}
	S(\gamma_{e,1},\gamma_{f,2},\gamma_{f,1},w)=-\left[(\gamma_{f,1}+\gamma_{f,2})^{2}-4w^{2}\right]^{1/2} + T(\gamma_{e,1},\gamma_{f,2},\gamma_{f,1},w) + T(\gamma_{e,1},\gamma_{f,1},\gamma_{f,2},w)
\end{equation}
where 
\begin{equation}
	\begin{aligned}
	T(x,y,z,w) =& w^{3}(y z)^{-3/2}(y z - 1)h^{-1}\left[A_{0}(h)-(1+h)^{1/2}\right]-(1+h)^{1/2}w^{-1}(y z)^{-1/2} \\ 
				&+ (w/2)(y z)^{-3/2}\left[(1+h)^{-1/2}(y^{2}+y z +x z - x y -2w^{2})-4 y z A_{0}(h) \right]\\
	\end{aligned}
\end{equation}
where
\begin{equation}
	A_{0}(h)
		\begin{cases}
			= h^{-1/2}\ln\left[h^{1/2}+(1+h)^{1/2}\right] & {\rm ~ for ~} h>0 \\
			=(-h)^{-1/2}arcsin\sqrt{-h} & {\rm ~ for ~} h<0 \\
			\approx 1-h/6+3h^{2}/40 & {\rm ~ for ~} h\approx0 \\
		\end{cases}
\end{equation}
and finally, with $h=\left[(x-y)^{2}-1\right]w^{2}/y z$ and the integration boundaries $w_{L}=w_{-}$, $w_{U}=min\left[\sqrt{\gamma_{f,1}\gamma_{f,2}},w_{+}\right]$, in which, 
\begin{equation}
	w_{\pm}=\left[ \gamma_{e,1}\gamma_{e,2} + 1 \pm \sqrt{(\gamma_{e,1}^{2}+1) (\gamma_{e,2}^{2}+1)} \right]/2 \ .
\end{equation}

The emergence rate for photons, due to pair annihilation process, is 
\begin{equation}
	Q_\mathrm{pa}(\gamma_{f,1})=c\int_{\gamma_{e,2}^{*}}^{\infty}n(\gamma_{e,2})d\gamma_{e,2} \int_{\gamma_{e,1}^{*}}^{\infty}n(\gamma_{e,1})R_{pa}(\gamma_{f,1}\leftarrow\gamma_{e,1},\gamma_{e,2})
	\label{equ:Qpa}
\end{equation}

The expression of $R_{pa}$ can be obtained via the symmetry 
\begin{equation}
	R_\mathrm{pa}(\gamma_{f,1}\leftarrow\gamma_{e,1},\gamma_{e,2})=R_\mathrm{pa}(\gamma_{f,2}\leftarrow\gamma_{e,1},\gamma_{e,2})=R_{pp}(\gamma_{e,1}\leftarrow\gamma_{f,1},\gamma_{f,2})=R_\mathrm{pp}(\gamma_{e,2}\leftarrow\gamma_{f,1},\gamma_{f,2})
\end{equation}
and energy conservation 
\begin{equation}
\gamma_{f,1}+\gamma_{f,2}=\gamma_{e,1}+\gamma_{e,2}.
\end{equation}

The lower-limits of the integration in \equ{Qpp} are
\begin{equation}
	\begin{aligned}
		\gamma_{f,1}^{*} & =\dfrac{1}{2}\gamma_{e,1}(1-\beta_{e,1}) \\ 
		\gamma_{f,2}^{*} & = \begin{cases}
								\gamma_{f,2}/\{[2\gamma_{f,2}-\gamma_{e,1}(1+\beta_{e,1})]\gamma_{e,1}(1+\beta_{e,1})\} & {\rm~for~} x>x_{+}\\
								\gamma_{f,2}/\{[2\gamma_{f,2}-\gamma_{e,1}(1-\beta_{e,1})]\gamma_{e,1}(1-\beta_{e,1})\} & {\rm~for~} x<x_{-}\\
								\gamma_{e,1}-\gamma_{f,2}+1 & {\rm~for~} x_{-}\le x \le x_{+}\\
							\end{cases}
	\end{aligned}
\end{equation}
with $x_{\pm}=[1+\gamma_{e,1}(1\pm\beta_{e,1})]/2$ while the limits in \equ{Qpa} are
\begin{equation}
	\begin{aligned}
		\gamma_{e,2}^{*} = 
			\begin{cases}
				\gamma_{A} & {\rm ~for~} \gamma_{f,2}<1/2\\
				1 & {\rm ~for~} \gamma_{f,2}\ge 1/2\\
			\end{cases} \hspace{5mm}
		\gamma_{e,1}^{*} =
			\begin{cases}
				\gamma_{-} & {\rm ~for~} \gamma_{f,2}\le1/2\\
				\gamma_{-} & {\rm ~for~} 1/2<\gamma_{f,2}<1 {\rm ~and~} \gamma_{+}<\gamma_{B}\\
				\gamma_{+} & {\rm ~for~} \gamma_{f,2}\ge1 {\rm ~and~} \gamma_{+}<\gamma_{B}\\
				1 & {\rm~otherwise}\\
			\end{cases}
	\end{aligned}
\end{equation}
where 
\begin{equation}
	\gamma_{\pm}=(F_{\pm}+F_{\pm}^{-1})/2,\hspace{5mm}
	F_{\pm}=2\gamma_{f,2}-\gamma_{e,2}(1\pm\beta_{e,2}),\hspace{5mm}
	\gamma_{A}=\gamma_{f,2}+1/(4\gamma_{f,2}),\hspace{5mm}
	\gamma_{B}=\gamma_{f,2}-(\gamma_{f,2}-1)/(2\gamma_{f,2}-1)
\end{equation}

The disappearance rate for photon $\gamma_{f,2}$ due to pair production with target photon $\gamma_{f,1}$ is
\begin{equation}
	\alpha_\mathrm{pp}(\gamma_{f,2})=\int R_{pp}(\gamma_{f,2},\gamma_{f,1})n(\gamma_{f,1})d\gamma_{f,1}
\end{equation}
and $R_{pp}$ (in the unit of $\sigma_{T}$) is given by (see also, DM09)
\begin{equation}
	R(\gamma_{f,2},\gamma_{f,1})=\dfrac{3}{8}c\gamma_{f,2}^{-2}\gamma_{f,1}^{-2}\bar{\varphi}(u,v)
\end{equation}
and 
\begin{equation}
	\bar{\varphi}(u,v)=\left[2v+(1+v)^{-1}\right]\ln u - \ln^{2}u-2(2v+1)v^{1/2}(v+1)^{-1/2}+4\ln u \ln(1+u) +\pi^{2}/3+4{\rm Li}_{2}(-u)
	\label{equ:phibaruv}
\end{equation}
where 
\begin{equation}
	v=\gamma_{f,2}\gamma_{f,1}-1, \hspace{5mm} u=\dfrac{\sqrt{v+1}+\sqrt{v}}{\sqrt{v+1}-\sqrt{v}}
\end{equation}

The asymptotic form of \equ{phibaruv} is 
\begin{equation}
	\bar{\varphi}(u,v)\approx\begin{cases}
		2v\ln(4v)-4v+\ln^{2}(4v) & {\rm ~for~} v>>1 \\
		\dfrac{4}{3}v^{3/2} & {\rm ~for~} v<<1
		\end{cases}\ .
\end{equation}

The pair annihilation rate is
\begin{equation}
	\alpha_\mathrm{pa}(\gamma_{e,2})=\int R_\mathrm{pa}(\gamma_{e,1},\gamma_{e,2})n(\gamma_{e,1})d\gamma_{e,1}
\end{equation}
where the kernel, in the unit of $\sigma_{T}$, is
\begin{equation}
	R_\mathrm{pa}(\gamma_{e,1},\gamma_{e,2})=\dfrac{3}{8}c \gamma_{e,1}^{-2}\gamma_{e,2}^{-2}\left[S_\mathrm{pa}(\gamma_{e}^{+}) - S_\mathrm{pa}(\gamma_{e}^{-})\right]\ ,
\end{equation}
\begin{equation}
	S_\mathrm{pa}(x)=x^{2}\ln(4x^{2})-2x^{2}+\dfrac{3}{4}\ln^{2}(4x^{2})
\end{equation}
and 
\begin{equation}
	\gamma_{e}^{\pm}=\dfrac{1}{2}\left[\gamma_{e,1}\gamma_{e,2}+1\pm\sqrt{(\gamma_{e,1}^{2}-1)(\gamma_{e,2}^{2}-1)}\right]\ .
\end{equation}


\subsection{Synchrotron}


The emission rate of synchrotron photons by a relativistic electron $\gamma_{e}\gtrsim10$ is given by 

\begin{equation}
	Q_\mathrm{syn}(\gamma_{f})=\int R_{syn}(\gamma_{f}\leftarrow\gamma_{e})n(\gamma_{e})d\gamma_{e}\ .
\end{equation}

The integration kernel is well known as
\begin{equation}
	R_\mathrm{syn}(\gamma_{f}\leftarrow\gamma_{e})=c\sigma_{T}(3\sqrt{3}/\pi)\gamma_{f}^{-1}u_{b}F_{b}^{-1}z^{2}\left\{K_{4/3}(z)K_{1/3}(z)-(3/5)z\left[K_{4/3}^{2}(z)-K_{1/3}^{2}(z)\right]\right\}\,
	\label{equ:Rsyn}
\end{equation}
where 
\begin{equation}
	u_{b}=B^{2}/(8\pi m_{e}c^{2}),\hspace{5mm} F_{b}=B/B_\mathrm{crit},\hspace{5mm} B_\mathrm{crit}=(m_{e}c^{2})^{2}/(ce\hbar)=4.41\times10^{13}{\rm~G},\hspace{5mm} z=\gamma_{f}\gamma_{e}^{-2}F_{b}^{-1}/3 
\end{equation}
and $K_{n}(z)$ is the modified Bessel function of the second kind.

For the extinction rate for the photons due to synchrotron-self absorption effect, we follow the treatment of VP09:
\begin{equation}
	\alpha_\mathrm{ssa}(\gamma_{f})=\dfrac{\lambda_{C}^{3}}{8\pi}\int \gamma_{f}^{-1}R_\mathrm{syn}(\gamma_{f}\leftarrow\gamma_{e})\gamma_{e}^{2}\partial_{\gamma}\left[\gamma_{e}^{-2}n(\gamma_{e})\right] d\gamma_{e}
\end{equation}

where $\lambda_{C}=h/m_{e}c$ is the Compton wavelength. The cooling effect on electron, positron and protons, due to synchrotron and synchrotron-self absorption effects, is modeled as a continuous energy loss process and thus described by the differential terms 
\begin{equation}
	\dot{\gamma}_\mathrm{syn}(\gamma)=\dot{\gamma}_{s}(\gamma)+2\gamma^{-1}{H}_{s}(\gamma)+\partial_{\gamma}{H}_{s}(\gamma), \hspace{5mm} D_\mathrm{syn}(\gamma)=2{H}_{s}(\gamma)
	\label{equ:gammadotsyn}
\end{equation}
and
\begin{equation}
	\dot{\gamma}_{s}=-\dfrac{4}{3}\sigma_{T}u_{B}\gamma_{e}^{2}\,
\end{equation}
\begin{equation}
	H_{s}(\gamma_e)=\dfrac{\lambda_{C}^{3}}{8\pi}\int R_\mathrm{syn}(\gamma_{f}\leftarrow\gamma_{e})n(\gamma_{f})d\gamma_{f}
	\label{equ:Hsyn}
\end{equation}


\subsection{$p\gamma$ interaction}


For $p\gamma$ interaction $p+f\rightarrow X$, here with $p$ being a proton, $f$ the target photon and $X$ representing a proton, neutron or pion, we follow the simplified treatment {\it sim A} in H10, and incorporate them in this numerical framework. The generation rate of particle $X$ can be expressed as
\begin{equation} 
	Q_{X}(E_{X})=\int n_{f}(E_{f})R_{p\gamma}(E_{X}\leftarrow E_{f})dE_{f}
\end{equation}
where $Q\equiv d\dot{n}_{X}/dE_{X}$ and $n\equiv d^{2}N/dVdE$ so that the phenomenological values in the references can be directly applied. 

The response function $R_{p\gamma}(E_{X}\leftarrow E_{f})$ depends on a further layer of integration:
\begin{equation}
	R_{p\gamma}(E_{X})=\int R_{p\gamma}(E_{X}\leftarrow E_{p},E_{f})n_{p}(E_{p})dE_{p}
\end{equation}
where the integration kernel is 
\begin{equation}
	R_{p\gamma}(E_{X}\leftarrow E_{p},E_{f})=cE_{p}^{-1}\sum_{i}\delta(x-\chi_{i})M_{i}f_{i}(y)
\end{equation}
where $i$ stands for the interaction channel, $M_{i}$ is the multiplicity of the particle $X$, $\chi_{i}$ is the inelasticity of the collision, $f(y)$ is defined as an integration over the cross-section as a function of photon energy $\epsilon_{r}$ expressed in the particle rest-frame:
\begin{equation}
	f(y)=\dfrac{1}{2y^{2}}\int_{\epsilon_{th}}^{2y} d\epsilon_{r}\epsilon_{r}\sigma(\epsilon_{r})
\end{equation}
and finally, with $x= E_{X}/E_{p}$ and $y=\gamma_{p}E_{f}$.

Under the $\delta-$function approximation, $R_{p\gamma}(E_{X}\leftarrow E_{f})$ simplifies to
\begin{equation}
	R_{p\gamma}(E_{X}\leftarrow E_{f})=c\sum_{i}E_{p,0}^{-1} \chi_{i}^{-2}E_{X}M_{i}f_{i}(y_{0})n_{p}(E_{p,0})
\end{equation}
where $E_{p,0}$ and $y_{0}$ corresponds to the solution of $x-\chi_{i}=0$. The expression of $Q_{X}$ simplifies to 
\begin{equation}
	Q_{X}(E_{X})=c n_{p}(\chi_{i}^{-1}E_{X})E_{X}^{-1}m_{p}\sum_{i}M_{i}\int n_{f}(E_{X}^{-1}\chi_{i}m_{p}y)f_{i}(y)dy
	\label{equ:QxEx}
\end{equation}

The disappearance rate, due to participation in $p\gamma$ process for protons is 
\begin{equation}
	\begin{aligned}
		\alpha_{p\gamma}(E_{p})	= & \int dE_{X} R_{p\gamma}(E_{X}\leftarrow E_{p}) = \int dE_{X} \int dE_{f}R_{p\gamma}(E_{X}\leftarrow E_{p},E_{f})n_{f}(E_{f})\\
								= & c \sum_{i}M_{i}\int dE_{f}n_{f}(E_{f})f_{i}(\gamma_{p}E_{f})
	\end{aligned}
\end{equation}
and for photon
\begin{equation}
	\alpha_{p\gamma}(E_{f}) = \int dE_{X}R_{p\gamma}(E_{X}\leftarrow E_{f})=c\sum_{i}\chi_{i}^{-1}M_{i}\int dE_{X}f_{i}(\chi_{i}^{-1}E_{X}E_{f}m_{p}^{-1})n_{p}(\chi_{i}^{-1}E_{X})
\end{equation}

The expressions of $f_{i}(y)$ and coefficients are given by eqn.30,33-35,40 and Table.3,5,6 of (ref. HU10) when $i$ falls into the category of resonance, direct and multi-pion production, respectively.


\subsection{Bethe-Heitler}


For the Bethe-Heitler or photopair process $p+\gamma_{f}\rightarrow \gamma_{e,1}+\gamma_{e,2}$, we incorporate the calculation under the framework similar to the multi-pion production in the previous section. The generation rate for pairs are given by \equ{QxEx} where the multiplicity $M=1$ for $e^{-}$ and $e^{+}$, respectively. The cross-section and inelasticity are approximated by a series of step-functions, where each step is defined as an interaction-channel $i$. The cross-section $\sigma_{\!_\mathrm{BH}}(\gamma_{f}^{\prime})$ as a function of photon energy (in the unit of $m_{e}c^{2}$) $\gamma_{f}^{\prime}$ in the proton rest frame, obtained under the approximation of zero recoil of the proton, can be expressed as
\begin{equation}
	\sigma_{\!_\mathrm{BH,tot}}(\gamma_{f}^{\prime})=\int_{1}^{\gamma_{f}^{\prime}}d\gamma_{e,1}\int_{-1}^{+1}d\mu_{e,1}\dfrac{d^{2}\sigma_{\!_\mathrm{BH,diff}}}{d\gamma_{e,1}d\mu_{e,1}}
	\label{equ:sigmaBHt}
\end{equation}
where the differential cross-section in the proton rest frame is expressed as \citep{Blumenthal1970}:
\begin{equation}
	\begin{aligned}
		\dfrac{d^{2}\sigma_{\!_\mathrm{BH,diff}}}{d\gamma_{e,1}d\mu_{e,1}} = & 
			\left(\dfrac{3\alpha_{f}\sigma_{T}p_{1}p_{2}}{16\pi\gamma_{f}^{3}}\right)  \left[-4(1-\mu^{2})\dfrac{2\gamma_{1}^{2}+1}{p_{1}^{2}\Delta_{1}^{4}} +\dfrac{5\gamma_{1}^{2}-2\gamma_{1}\gamma_{2}+3}{p_{1}^{2}\Delta_{1}^{2}} + \dfrac{p_{1}^{2}-\gamma_{f}^{2}}{T^{2}\Delta_{1}^{2}} + \dfrac{2\gamma_{2}}{p_{1}^{2}\Delta_{1}} + \dfrac{Y}{p_{1}p_{2}} \right. \left( 2\gamma_{1}(1-\mu^{2})\dfrac{3\gamma_{f}+p_{1}^{2}\gamma_{2}}{\Delta_{1}^{4}} \right.\\ 
			 & \left. \left. + \dfrac{2\gamma_{1}^{2}(\gamma_{1}^{2}+\gamma_{2}^{2})-7\gamma_{1}^{2}-3\gamma_{1}\gamma_{2}-\gamma_{2}^{2}+1}{\Delta_{1}^{2}} + \dfrac{\gamma_{f}(\gamma_{1}^{2}-\gamma_{1}\gamma_{2}-1)}{\Delta_{1}} \right) -\dfrac{\delta_{+}^{T}}{p_{2}T} \left( \dfrac{2}{\Delta_{1}^{2}} - \dfrac{3\gamma_{f}}{\Delta_{1}} - \dfrac{\gamma_{f}(p_{1}^{2}-\gamma_{f}^{2})}{T^{2}\Delta_{1}} \right) -\dfrac{2y_{+}}{\Delta_{1}} \right]\,
	\end{aligned}
\end{equation}
where
\begin{equation}
	T=| \mathbf{k}- \mathbf{p}_{1}|,\hspace{5mm} Y=\dfrac{2}{p_{1}^{2}}\ln\left[\dfrac{\gamma_{1}\gamma_{2}+p_{1}p_{2}+1}{\gamma_{f}}\right],\hspace{5mm}
	y_{+}=p_{2}^{-1}\ln\left[\dfrac{\gamma_{2}+p_{2}}{\gamma_{2}-p_{2}}\right],\hspace{5mm} \delta_{+}^{T}=\ln\left[\dfrac{T+p_{2}}{T-p_{2}}\right],\hspace{5mm}
	\Delta_{1}=\gamma_{1}(1-\beta_{1}\mu)\,
\end{equation}
and in the above equations we have dropped the primes and the electron sub-index $e$ (in which, $1=e^{-},2=e^{+}$).

The inelasticity for one electron can be obtained via
\begin{equation}
	\chi_{\!_\mathrm{BH}}(\gamma_{f})=\dfrac{m_{e}}{m_{p}}\int_{1}^{\gamma_{f}-1}d\gamma_{1}\int_{-1}^{+1}d\mu \Delta_{1}\dfrac{d^{2}\sigma_{\!_\mathrm{BH,diff}}}{d\gamma_{1}d\mu}\ .
	\label{equ:chiBH}
\end{equation}

\eqs~(\ref{equ:sigmaBHt}) and (\ref{equ:chiBH} are approximated as a sum of step functions:
\begin{equation}
	\begin{aligned}
		\sigma_{\!_\mathrm{BH,tot}}(\gamma_{f}) & = \sum_{i} \sigma_{\!_{\mathrm{BH},i,0}} H(\gamma_{f}-\gamma_{f,i}^\mathrm{min})H(\gamma_{f,i}^\mathrm{max}-\gamma_{f}) \\
		\chi_{\!_\mathrm{BH}}(\gamma_{f}) & = \sum_{i}\chi_{\!_{\mathrm{BH},i,0}} H(\gamma_{f}-\gamma_{f,i}^\mathrm{min})H(\gamma_{f,i}^\mathrm{max}-\gamma_{f})\\
	\end{aligned}
\end{equation}
where
\begin{equation}
	\begin{aligned}
		\sigma_{\!_{\mathrm{BH},i,0}}(\gamma_{f})& = \dfrac{1}{\gamma_{f,i}^\mathrm{max}-\gamma_{f,i}^\mathrm{min}}\int_{\gamma_{f,i}^\mathrm{min}}^{\gamma_{f,i}^\mathrm{max}}d\gamma_{f}\sigma_{\!_\mathrm{BH,tot}}(\gamma_{f})\\
		\chi_{\!_{\mathrm{BH},i,0}}(\gamma_{f}) & = \dfrac{1}{\gamma_{f,i}^\mathrm{max}-\gamma_{f,i}^\mathrm{min}}\int_{\gamma_{f,i}^\mathrm{min}}^{\gamma_{f,i}^\mathrm{max}}d\gamma_{f}\chi_{\!_\mathrm{BH}}(\gamma_{f})\, \\
	\end{aligned}
\end{equation}
and the functions \{$f_{\!_{\mathrm{BH},i}}(y)$\} are obtained by eqn.40 of HU10 by replacing the coefficients with the ones from the above equation.

The extinction function on photons are
\begin{equation}
	\alpha_{\!_\mathrm{BH}}(\gamma_{f})=2c\sum_{i}\chi_{\!_{\mathrm{BH}},i,0}^{-1}\int dE_{1}f_{\!_{\mathrm{BH},i}}(\chi_{\!_{\mathrm{BH},i,0}}^{-1}E_{1}E_{f}m_{p}^{-1})n_{p}(\chi_{\!_{\mathrm{BH},i,0}}^{-1}E_{1})\ .
\end{equation}

For protons, the generation rate in principle can be obtained from \equ{QxEx} by substituting $M_{i}=1$, $f_{i}=f_{\!_{\mathrm{BH},i}}$ and $\chi_{i}=1-2\chi_{\!_{\mathrm{BH},i,0}}$. However, since $\chi_{\!_{\mathrm{BH},i,0}}\ll1$, $\chi_{i}\approx1$, which the term $n_{p}(\chi_{i}^{-1}E_{p})$ almost overlaps with the parent proton bin $n_{p}(E_{p})$ and the numerical grid cannot resolve this difference. Instead, similar as the case of inverse Compton scattering in the Thomson regime, this effect is treated as a continuous energy loss process and thus described by the differential terms. By requiring the equality between the integral terms and the differential terms, such as \equ{continuousloss}, we have
\begin{equation}
	\begin{aligned}
		\dot{E}_{\!_\mathrm{BH}}(E_{p})&= \sum_{i}-2\chi_{\!_{\mathrm{BH},i,0}}E_{p}\alpha_{p,i}(E_{p})\\
		D_{\!_\mathrm{BH}}(E_{p}) & = \sum_{i}4\chi_{\!_\mathrm{BH}}^{2}E_{p}^{2}\alpha_{p,i}(E_{p})\, \\
	\end{aligned}
	\label{equ:dbh}
\end{equation}
where we have defined $\alpha_{p,i}(E_{p})=2c\int dE_{f}n_{f}(E_{f})f_{\!_{\mathrm{BH},i}}(\gamma_{p}E_{f})$.

\subsection{$\pi$ decay kinematics}

The dominant decay channels of $\pi^{\pm}$ and $\mu^{\pm}$ are

\begin{equation}
	\begin{aligned}
		\pi^{+}(\pi^{-}) & \rightarrow \mu^{+}(\mu^{-})+\nu_{\mu}(\bar{\nu}_\mu) \\
		\mu^{+}(\mu^{-}) & \rightarrow e^{+}(e^{-})+\nu_{e}(\bar{\nu}_{e})+\bar{\nu}_{\mu}(\nu_{\mu})
	\end{aligned}
\end{equation}

The probability density function for the decay product $j$ from the parent particle $i$, as a function of scaling variable $x=E_{j}/E_{i}$, is
\citep{Lipari07}
\begin{equation}
	\begin{aligned}
		f_{\mbox{$\mu$}_{\mbox{\scriptsize{$R$}}}^{\mbox{\small {$+$} } } }(x) =
		f_{\mbox{$\mu$}_{\mbox{\scriptsize{$L$}}}^{\mbox{\small {$-$} } } }(x) =
		\dfrac{r_{\mu\pi}^{2}(1-x)}{(1-r_{\mu\pi}^{2})^{2}x}H(x-r_{\mu\pi}^{2}) \\
		f_{\mbox{$\mu$}_{\mbox{\scriptsize{$L$}}}^{\mbox{\small {$+$} } } }(x) =
		f_{\mbox{$\mu$}_{\mbox{\scriptsize{$R$}}}^{\mbox{\small {$-$} } } }(x) =
		\dfrac{x-r_{\mu\pi}^{2}}{(1-r_{\mu\pi}^{2})^{2}x}H(x-r_{\mu\pi}^{2})
	\end{aligned}
\end{equation}

Due to energy conservation, the corresponding neutrino directly from pion decay has a distribution function of

\begin{equation}
	f_{\nu}(1-x)=f_{\mu}(x)
\end{equation}

For $\pi^{0}\rightarrow\gamma+\gamma$, the distribution function for a photon is

\begin{equation}
	f_{\gamma}(x)=\dfrac{1}{2\beta_{\pi}\gamma_{\pi}}H\left(x-\dfrac{1-\beta_{\pi}}{2}\right)H\left(\dfrac{1+\beta_{\pi}}{2}-x\right)\ .
\end{equation}

The neutrino distribution function from muon decay is
\begin{equation}
	\begin{aligned}
		f_{\mbox{$\bar{\nu}$}_{\mbox{\scriptsize{$\mu$}}} }(x,h) &=
		f_{\mbox{$\nu$}_{\mbox{\scriptsize{$\mu$}}} }(x,-h) =
		\left(\dfrac{5}{3}-3x^{2}+\dfrac{4}{3}x^{3}\right)+h\times\left(-\dfrac{1}{3}+3x^{2}-\dfrac{8}{3}x^{3}\right)\\
		f_{\mbox{$\nu$}_{\mbox{\scriptsize{$e$}}} }(x,h) &=
		f_{\mbox{$\bar{\nu}$}_{\mbox{\scriptsize{$e$}}} }(x,-h) =
		(2-6x^{2}+4x^{3})+h\times(2-12x+18x^{2}-8x^{3})
	\end{aligned}
\end{equation}
and for $e^{\pm}$, since we no longer need to distinguish between their chiralities in this paper, the distribution can be simply expressed as (e.g. Particle Data Group)
\begin{equation}
	f_{e}(x)=\dfrac{4}{3}(1-x^{3})
\end{equation}
under the relativistic approximation, $\gamma_{e}\gtrsim10$.


\section{numerical treatment}
\label{appendix:num}

On the numerical aspect, it is more convenient to rewrite \equ{overalllinearform} in the form where energy is expressed on the logarithmic scale
\begin{equation}
	\partial_{t}n(x,t)=-\partial_{x}\left[A(x,t)n(x,t)-B(x,t)\partial_{x}n(x,t)\right]-\alpha(x,t)n(x,t)+\epsilon(x,t)
	\label{equ:overalllogarithmicform}
\end{equation}
by making the sustitutes 
\begin{equation}
	x=\ln\gamma, \hspace{5mm} 
	n(x)=\gamma n(\gamma), \hspace{5mm}
	A(x)=\dfrac{\dot{\gamma}}{\gamma}-\partial_{\gamma}\left[\dfrac{D(\gamma)}{2\gamma}\right], \hspace{5mm}
	B(x)=\dfrac{D(\gamma)}{2\gamma}, \hspace{5mm}
	\epsilon(x)=\gamma Q(\gamma), \hspace{5mm}
	\alpha(x)=\alpha(\gamma)\, 
\end{equation}
where the terms on the right-hand sides are the ones from \equ{overalllinearform} while the left-hand sides correspond to \equ{overalllogarithmicform}.

The x-axis is equally spaced by width of $\Delta{x}$ from $x_{min}$ to $x_{max}$, which represents energy on logarithmic scale
\begin{equation}
	n_{i}=n_\mathrm{min}+(i-1)\Delta{x}, \hspace{5mm}i=1,2,...i_{max}
\end{equation}
and t-axis is linear in time, and equally spaced by width of $\Delta{t}$
\begin{equation}
	n^{k}=n^{0}+(k-1)\Delta{t}, \hspace{5mm}k=1,2,...t_{max}\ .
\end{equation}

The discrete form of \equ{overalllogarithmicform} can be written, with the abbreviated notations here on the index of any quantity $s$:
\begin{equation}
s\equiv s_{i}^{k},~s_{1}\equiv s_{i+1},~s_{-1}\equiv s_{i-1},~s^{1}\equiv s^{k+1}, etc.
\end{equation}
as
\begin{equation}
	(n^{1}-n)/\Delta{t}=-(F_{+}^{+}-F_{-}^{+})/\Delta{x}-\alpha \times(n^{1}-n)+\epsilon
\end{equation}
where
\begin{equation}
	F_{\pm}^{+}=A_{\pm}n_{\pm}^{+}-B_{\pm}(n_{\pm}^{+}-n^{+})/\Delta{x}\ .
\end{equation}

The differential scheme on time is Crank-Nicolson, which is
\begin{equation}
	n^{+}=(n^{1}+n)/2
\end{equation}
and on energy is Chang \& Cooper \citep{Chang1970}, which is 
\begin{equation}
	n_{+}=(1-\delta)n_{1}+\delta*{n}
\end{equation}
where 
\begin{equation}
	\delta=\dfrac{1}{w}-\dfrac{1}{e^{w}-1}, \hspace{5mm}w=-\dfrac{A_{+}}{B_{+}}\Delta{x}\ .
\end{equation}

At very high energies, the secondary electron or positron lose energy rapidly within a feasible choice of computational timestep $\Delta t$ and this process cannot be properly calculated from the above framework. Instead, in this region, we switch to a semi-analytical approach of \equ{overalllogarithmicform} but have neglected the second-order differential term $B(x)$, which is a trade-off between a small fraction of accuracy and orders-of-magnitude increase in efficiency. Since in this region the cooling coefficient $A(x)$ is mainly contributed by low-energy target-photons and the magnetic field, with a proper choice of $\Delta t$ and partition of this region, 1) $A(x)$ can be treated as a constant within each $\Delta t$; 2) the time-dependent solution in each energy bin quickly converges to a steady-state solution within $\Delta t$. Therefore, semi-analytical solutions can be obtained and represented for the electron and positron population for each time-step $\Delta t$, as the following
\begin{equation}
	n(x,t+\Delta t)=\dfrac{A(y,t)}{A(x,t)}exp\left[-\int_{y}^{x}\dfrac{\alpha(x^{\prime},t)}{A(x^{\prime},t)}dx^{\prime}\right]n(y,t)+\dfrac{1}{A(x,t)}\int_{y}^{x} dx^{\prime}\epsilon(x^{\prime},t)exp\left[-\int_{x^{\prime}}^{x}\dfrac{\alpha(x^{\prime\prime},t)}{A(x^{\prime\prime},t)}dx^{\prime\prime}\right]
\end{equation}
where $y$ is the solution of the equation
\begin{equation}
	-\int_{1}^{y}\dfrac{dx^{\prime}}{A(x^{\prime},t)}=-\int_{1}^{x}\dfrac{dx^{\prime}}{A(x^{\prime},t)}+\Delta{t}\, 
\end{equation}
which can be easily calculated numerically.

\end{document}